\documentclass[
    twocolumn,
	prd,
	amssymb,
	preprintnumbers,
	secnumarabic,
	nofootinbib]{revtex4-1}

\pdfoutput=1

\usepackage{bigints}
\usepackage{bm}
\usepackage{tikz}
\usepackage{graphicx}
\usepackage{multirow}
\usepackage{amsmath}
\usepackage{amssymb}
\usepackage{hyperref}
\usepackage{color}
\usepackage{xspace}
\usepackage{verbatim}
\usepackage{mathtools}
\usepackage{booktabs}
\usepackage{graphicx}
\usepackage{dcolumn}
\usepackage{bm}
\usepackage{hyperref}

\definecolor{aquamarine}{rgb}{0.2,0.7,0.6}
\definecolor{cerulean}{RGB}{0,166,214} 
\definecolor{subtlered}{rgb}{0.8,0.3,0.3}

\begin{document}

\title{Ultraheavy multiscattering dark matter:\\ DUNE, CYGNUS, kilotonne detectors, and tidal streams}

\author{Harsh Aggarwal}
 \email{harsh2002@kgpian.iitkgp.ac.in}
 \affiliation{Department of Physics, Indian Institute of Technology Kharagpur, Kharagpur 721302, India}
\author{Nirmal Raj}%
 \email{nraj@iisc.ac.in}
\affiliation{%
 Centre for High Energy Physics, Indian Institute of Science, C. V. Raman Avenue, Bengaluru 560012, India
}%

\begin{abstract}
In direct searches of dark matter multi-scatter signatures are now being sought to probe scattering cross sections large enough to make the detector optically thick to incident particles.
We provide some significant updates to the multi-scatter program.
Using considerations of energy deposition, we derive the reaches in cross section and mass of various proposed large-volume detectors: a kilotonne fiducial mass ``module of opportunity" at DUNE, 
a kilotonne xenon detector suggested for neutrinoless double beta decay,
the gaseous detector CYGNUS, and the dark matter detectors XLZD and Argo.
Where the velocity vector can be reconstructed event-by-event, the Galactic dark matter velocity distribution may be inferred. 
We exploit this to show that halo substructure such as tidal streams can be picked up if they make up about 10\% of the local dark matter density. 
\end{abstract}

\maketitle

\section{\label{sec:intro}Introduction}

Seeking direct detection of the flux of halo dark matter (DM) in laboratories underground has been a decades-long endeavor that has branched out into multiple new, exciting directions.
One such is the renewed interest in searches for
DM scattering multiple times during transit, constraining per-nuclear scattering cross sections comparable to that of QCD interactions.
These multi-scatter searches are most relevant for DM masses above about $10^{13}$~GeV~\cite{GoodmanWitten:1984dc}, where the DM flux (inversely proportional to the DM mass) is the smallest, thus requiring large cross sections to compensate.
These ultra-heavy DM states may be produced by several mechanisms in the early universe~\cite{snowmass:Carney:2022gse}, and could be 
WIMPzilla-like~\cite{Models:Chung1998wimpzillas,*Models:kolb1998wimpzillas,*Models:Harigaya2016:GUTzillas}, colored~\cite{Models:ColoredDM}, 
baryon-charged~\cite{Models:BNL:DarkBaryonGeVMediator}, 
composite~\cite{Models:nucleiHardy:2014mqa,*Models:nucleiHardy:2015boa,*Models:nucleiMonroe:2016hic,*Models:Nuggets},
elementary~\cite{Bramante:2018tos},
dark monopoles~\cite{Models:EWSymMonopoles:Bai:2020ttp}, electroweak-symmetric solitons~\cite{Models:ElectroweakBalls}, and Planck-scale black hole relics~\cite{Models:PlanckScaleBHRelicsBaiOrlofsky2019,*Models:PlanckScaleBHRelicsSantaCruz}.
On the experimental end, recent multiscatter searches have limited DM-nucleus sub-barn cross sections for up to the Planck mass ($10^{19}$~GeV) scale~\cite{DAMA1999,EdelweissCDMSAlbuquerqueBaudis2003,KavanaghSuperheavy:2017cru,PICO:MS:Broerman2022,XENON1T:MSSSprojexn:Clark:2020mna,PlasticEtch:Bhoonah2020fys,CollarBeacomCappiello2021, DEAP:MS:2021raj,XENON1T:MS:2023iku,LZ:MS:2024psa};
for a recent white paper on the topic, see Ref.~\cite{snowmass:Carney:2022gse}.
The constrained regions generally have some overlap with the highest cross sections reachable by traditional single-scatter ``WIMP" searches~\cite{Raj:2024guv}.

Multi-scatter signals enjoy some unique virtues that single-scatter signals don't.
From the scatter multiplicity, {\em i.e.} the number of scatters in transit, the cross section may be reconstructed;
from the angle of acceptance of tracks in the detector, the DM mass may be inferred; 
then from the number of events, which is simply the integrated flux (up to efficiencies in detecting nuclear recoils), the local DM density may be inferred~\cite{Bramante:2018qbc}.
The local DM density may also be inferred from terrestrial scattering~\cite{Kavanagh:2020cvn}.
Where localization of an individual recoil site is possible, the DM track itself can be reconstructed.
And if sensitive timing information is available, the DM velocity vector may also be reconstructed event-by-event.
This would in turn help to directly measure parameters of the Galactic halo velocity distribution, determine any anisotropies of the local distribution in the Galactic frame, and possibly distinguish DM signals from some backgrounds~\cite{Bramante:2019yss}.
This style of performing astrometry of the DM halo can be compared with directional detection, where the directions of target particle recoils are used to infer the directionality of the DM flux~\cite{reviewDirectional:Vahsen:2021gnb}. 

In this work we revisit multi-scatter DM and provide significant updates.
Earlier work~\cite{Bramante:2018tos,Bramante:2019yss} had identified large-volume liquid scintillator-based neutrino detectors such as the currently operational SNO+ and soon-to-come JUNO as suitable for performing multiscatter searches of ultra-heavy dark matter.
The basic idea was that if enough energy from nuclear recoils is deposited in a short enough transit time, an excess over the inherent noise in photomultiplier tubes (PMTs) may be detected.
It was also argued that Cerenkov light-based detectors like SNO and Super-K, and charge readout-based detectors like DUNE, were unsuitable for this search as their energy thresholds were too high~\cite{Bramante:2018tos,Bramante:2019yss}. 

We now revisit this in the context of multiple large-volume detectors proposed in the literature.
At the imminent liquid argon-based neutrino experiment DUNE, four 10 kt modules would make up the far detector.
While the first three modules, using a combination of single-phase and dual-phase argon, have charge readout thresholds of about 10 MeV that would make it impossible to detect $\mathcal{O}$(10) keV recoils, the design of DUNE's fourth ``module of opportunity" is still not finalized. 
On the strength of this, Ref.~\cite{DUNEModuleDM:PNL2020,*DUNEModuleDM:snowmass:Avasthi2022,*DUNEModuleDM:Bezerra2023} argued for making it an unprecedented kilotonne-mass dark matter detector.
This could be achieved with a highly fiducialized  low-radioactivity argon dual-phase detector with readout through silicon photomultipliers (SiPMs) or ARIADNE cameras~\cite{ARIADNE:Roberts:2018sww} with near-$4\pi$ coverage, which will provide for recoil energy thresholds of 50--100 keV. 
While Ref.~\cite{DUNEModuleDM:PNL2020,*DUNEModuleDM:snowmass:Avasthi2022,*DUNEModuleDM:Bezerra2023} focused on conventional single-scatter searches, we will exploit its design to estimate the sensitivities of multiscatter signatures, and also show that due to the large flux admitted DM masses of $10^{21}$ GeV may be reached.
We will also study the feasibility of discovering multi-scattering ultraheavy DM at a kiltonne fiducial mass liquid xenon detector proposed mainly for neutrinoless double beta decay ($0\nu\beta\beta$) experiments but also suggested for dark matter searches~\cite{Xekton:Avasthi:2021lgy,Xekton:Anker:2024xfz}; we call this detector ``Xe-1kT".
Next we do the same at the future CYGNUS experiment, whose main goal is to perform sensitive directional detection in a gaseous detector~\cite{CYGNUS:2020pzb}.
Due to the smaller density of the gas than noble liquids, CYGNUS could probe DM-nucleus cross sections larger than those reached by DUNE and Xe-1kT. 
Finally, we identify the multiscatter reaches of the future multi-ton DM detectors XLZD~\cite{DARWIN:Macolino:2020uqq,XLZD:2024gxx} (formerly DARWIN) and Argo~\cite{ARGO:2018,ARGOSnowmassLOI} more carefully than done in the literature.

We also consider a concrete realization of anisotropies in the DM velocity distribution, namely, tidal streams -- halo substructure not virialized with the host galaxy, for which observational evidence exists~\cite{streamSDSSGaia2017,streamGaias2018,streamGaia2020,streamGaiaDR32022,streamLAMOST2023} -- that is passing through the Earth. 
We construct statistical tests to determine the presence of streams containing DM using the empirically available DM velocity distribution at our detectors. 
We show that streams with phase space properties markedly different from the background halo could be picked up if they make up at least about 10\% of the local dark matter density.

This study is laid out as follows.
In Section~\ref{sec:crosssecvmass} we estimate the DM-nucleus scattering cross section and DM mass sensitivities of DUNE, Xe-1kT, CYGNUS, XLZD, and Argo to ultraheavy multiscattering DM, using detector capabilities described in the literature.
We also discuss DM velocity vector reconstruction capabilities.
In Section~\ref{sec:streams} we discuss the identification of tidal streams in our setup, perform a median direction hypothesis test, and estimate the statistical power of discerning some well-studied streams.
In Section~\ref{sec:concs} we provide more discussion and conclude.
In the appendix we outline the computation of DM flux through a general detector setup, and use it to estimate the flux for departures from spherical detectors.
Among other things, we give a new prescription for estimating the flux through a nearly spherical detector accounting for the interplay between detector geometry and the anisotropy of the velocity distribution in the lab frame.

\section{\label{sec:crosssecvmass}Dark matter sensitivities}

\subsection{General considerations}

For the sake of obtaining order-of-magnitude estimates of the reach of future multiscatter searches we take the detector geometry to be spherical.
In Fig.~\ref{fig:xsvmreaches} we show, using dashed curves, our estimates of the 10 year live-time reaches of future large-volume detectors to the DM-nucleon cross section $\sigma_{\chi N}$ and DM mass $m_\chi$. 
For comparison are also shown using solid curves existing limits from underground multiscatter searches~\cite{DEAP:MS:2021raj,XENON1T:MS:2023iku,LZ:MS:2024psa}.
Some of these regions overlap with single-scatter search limits, which we do not display here but which can be read from Refs.~\cite{Raj:2024guv,XENON1T:MS:2023iku,LZ:MS:2024psa}.

Our future sensitivities are obtained as follows. 
We define a multiscatter event as one in which at least two DM-nucleus scatters per transit are observed.
The total number of such events is given by 
\begin{align}
\begin{split}
\label{eq:eventms}
N^{\rm MS}_{\rm ev} = \Phi(1-p(\lambda,0)-p(\lambda,1))~,
\end{split}
\end{align}
where $\Phi$ is the integrated flux of DM through the detector:
\begin{equation}
    \Phi =
(\rho_{\chi}/m_\chi)\pi R_{\text{fid}}^2~\int_0^{t_{\rm exp}} dt \int d^3 u \, uf_{\rm lab}(\vec{u},t)~,
\label{eq:fluxsphere}
\end{equation}
where $\rho_\chi$ = 0.3 $\rm GeV/cm^3$ is the local DM density,
$R_{\rm fid}$ the radius of the fiducial volume,
$t_{\rm exp}$ is the total live-time, $f_{\rm lab}$ is the lab-frame velocity distribution that will be described shortly.
In the appendix we provide notes on how to obtain the flux for fiducial volumes that are {\em near}-spherical or cuboidal.

In Eq.~\eqref{eq:eventms} $p(\lambda, k)$ is the Poisson probability of observing a multiplicity ({\em i.e.}, number of scatters per transit) of $k$ for an expected multiplicity of $\lambda$:
\begin{align}
\begin{split}
\label{eq:pson}
p(\lambda, k)=\frac{\lambda^k e^{-\lambda}}{k!}~,
\end{split}
\end{align}
with
\begin{align}
\begin{split}
\label{eq:psonavg}
\lambda= \sigma_{\chi N}^{\rm eff}n_{T}L_{\rm avg}~,
\end{split}
\end{align}
where $n_{T}$ is the target nucleus density, $L_{\rm avg}= 4R_{\rm fid}/3$
is the average detector chord length, and $\sigma_{\chi N}^{\rm eff}$ is an effective cross section 
accounting for the finite range of recoil energies, form factor suppression, and integration of the rate over the velocity distribution~\cite{Raj:2024guv}:
\begin{equation}
 \sigma_{\chi N}^{\rm eff} =    \sigma_{\chi N} \frac{m_T}{2\mu_{T\chi}^2 \bar v} \int^{E_{\rm R, max}}_{E_{\rm R, min}} dE_R F^2(E_R) \eta(E_R)~.
\label{eq:sigmaTxeff}
\end{equation}
Here $\sigma_{\chi N}$ is the total DM-nucleus cross section,
$m_T$ is the target nuclide mass, $\mu_{T\chi}$ the DM-nuclide reduced mass,
$\bar{v}$ the average DM speed $\simeq 350$~km/s, 
$F(E_R)$ a nuclear form factor, and $\eta$ the usual velocity integral~\cite{Lewin:1995rx} (see Ref.~\cite{nufloor:GaspertGiampaMorrissey:2021gyj} for an analytic expression) $= \int_{v_{\rm min}} d^3 u f_{\rm lab}(\vec{u})/u$~, where $v_{\rm min} = \sqrt{m_T E_R/2\mu_{T\chi}^2}$.
For the distribution of velocities in the virialized Milky Way halo, we will adapt the Standard Halo model~\cite{Lewin:1995rx,Baxter:2021pqo}, an isotropic Maxwell-Boltzmann distribution in the Galactic frame: 
\begin{align}
\begin{split}
\label{eq:velocitydistribution}
f_{\text{gal}}(\vec{v}) \propto \exp\left(-\frac{\left|\vec{v}\right|^2}{2 \sigma_v^2}\right) \Theta(v_\mathrm{esc} - \left|\vec{v}\right|)~,
\end{split}
\end{align}
with
\begin{align}
\begin{split}
\label{eq:veltrans}
\vec{v}= \vec{v}_0+ \vec{v}_{\rm lab}+ \vec{v}^\odot_{\rm pec}+ \vec{v}^\oplus_{\rm rev}(t)+ \vec{v}^\oplus_{\rm rot}(t)~,
\end{split}
\end{align}
where $\Theta(x)$ is the Heaviside step function that imposes a cut-off at the galactic escape speed, which we take as $v_{\rm esc}$ = 600 km/s.
The dispersion speed $\sigma_v=v_0/\sqrt{2}$, is given by the local standard of rest, $v_0=220$km/s.
These values are taken from Ref.~\cite{SMITHLEWIN1990} and are within uncertainties of the parameters recommended in Ref.~\cite{Baxter:2021pqo}. The normalization has been computed numerically.
With $\vec{v}_0$, 
the Sun's peculiar velocity $\vec{v}^\odot_{\rm pec}$, 
the time-varying motion of the Earth around the Sun $\vec{v}^\oplus_{\rm rev}$
and the daily rotation of the Earth $\vec{v}^\oplus_{\rm rot}$, we can rewrite the velocity distribution in the terrestrial laboratory frame $f_{\rm lab}$ using the Galilean transformation in Eq.~\eqref{eq:veltrans}.
We use the results of Refs.~\cite{PhysRevD.84.023516,Mayet_2016,KavanaghSuperheavy:2017cru} in implementing this transformation, going first from laboratory  to equatorial co-ordinates, and then to Galactic co-ordinates.

We will assume that scattering is spin-independent, so that $F(E_R)$ can be taken as the Helm form factor~\cite{Helm:1956zz,Lewin:1995rx}, and that there is no scaling with atomic number $A$ between per-nucleon and per-nucleus cross sections, which may arise from tightly bound composite DM
opaque to the nucleus with geometric zero-momentum cross sections~\cite{Digman:2019wdm}. 
Our results can be trivially extended to other interaction structures that give rise to, say, $A^4$ and isospin-violating scalings, spin-dependent scattering, and so on. 
Our 90\% C.L. sensitivities are obtained by assuming zero observed events, corresponding to $N_{\rm ev}^{\rm MS}$ = 2.44 expected events~\cite{Feldman:1997qc}.
This is a good assumption since multi-scatter DM searches are expected to be effectively background-free~\cite{Bramante:2018qbc,DEAP:MS:2021raj,XENON1T:MS:2023iku,LZ:MS:2024psa}.

\subsection{Reaches of future large-volume detectors}
\label{subsec:reachesxsvmx}

\begin{figure}
\includegraphics[width=0.9\hsize]{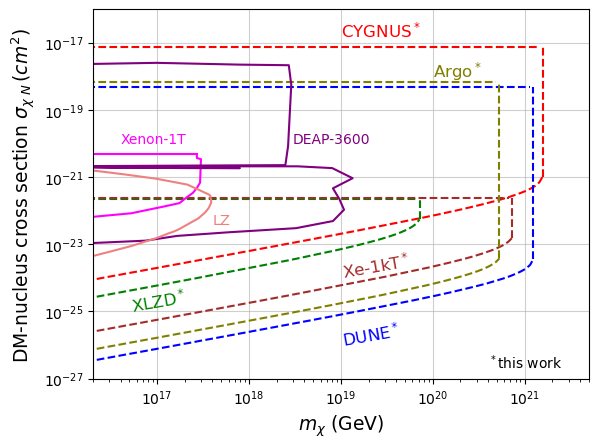}%
\caption{\label{fig:xsvmreaches} 
10 year live-time sensitivities of future large-volume detectors to DM-nucleus scattering cross sections and DM masses (dashed curves).
These correspond to no scaling between the per-nucleus and per-nucleon cross sections; other scenarios such as an $A^4$ scaling may be applied to these results.
The mass reach is set by the integrated flux in Eq.~\ref{eq:fluxsphere}, and the cross section sensitivities are set by considerations  discussed in Sec.~\ref{sec:crosssecvmass}.
Also shown with solid curves are limits from recent multiscatter searches at direct detection experiments.
See Sec.~\ref{sec:crosssecvmass} for further details.}
\end{figure}

To obtain the reach of DUNE, we assume as done in Ref.~\cite{DUNEModuleDM:PNL2020,*DUNEModuleDM:snowmass:Avasthi2022,*DUNEModuleDM:Bezerra2023} a module with 1~kt fiducial mass, which gives $R_{\rm fid}$ = 5.5~m for liquid argon density 1.4~g/cm$^3$, and an energy threshold of 75 keV per recoil.
In addition, we assume that a multiscatter search will proceed via a waveform analysis, making use of pulse shape discrimination (PSD) of nuclear vs electronic recoils, as done at the DEAP-3600 experiment~\cite{DEAP:MS:2021raj}.
(A multiscatter search using both PSD and the capabilities of a time-projection chamber (TPC) such as deployed at DarkSide-50 or liquid xenon-based experiments is another possibility, as explored in Ref.~\cite{DUNEModuleDM:PNL2020,*DUNEModuleDM:snowmass:Avasthi2022,*DUNEModuleDM:Bezerra2023}.)
We then obtain the reach on the cross section by simply requiring a multiplicity of at least 2, as just discussed. 
This results in a curve in Fig.~\ref{fig:xsvmreaches} with the exclusion cross section that scales logarithmically with $m_\chi$, which is roughly parallel to a similar curve obtained by the LZ multiscatter search, also shown in the figure. 
The maximum DM mass reached (again obtained by setting $N_{\rm ev}^{\rm MS}$ = 2.44) corresponds to running out of the DM flux $\propto m_\chi^{-1}$.
At DEAP-3600 the ceiling on the cross section came from the point at which the total energy deposition is so high that signal simulations were not computationally feasible.
The ceiling here is obtained by assuming a similar limitation, that the simulation that will be used for this analysis will break down for the same maximum number of scatters that DEAP-3600 could accommodate. 
This maximum multiplicity is obtained from the DEAP-3600 ceiling and fiducial volume. 

For Xe-1kT we take a fiducial mass of 1000 tonnes~\cite{Xekton:Avasthi:2021lgy,Xekton:Anker:2024xfz}, implying $R_{\rm fid}$ = 4.3 m for liquid xenon density of 2.94 g/cm$^3$, and assume a multi-scatter search similar to that performed by LZ~\cite{LZ:MS:2024psa}, but with threshold energy for individual recoils taken as 5 keV as suggested in the DARWIN proposal~\cite{DARWIN:Macolino:2020uqq}.
Once again the $\sigma_{\chi N}$ upper limit as a function of $m_\chi$ is parallel to the other curves due to Poisson fluctuations in multiplicity (Eq.~\eqref{eq:eventms}).

CYGNUS is a proposed gaseous TPC, which we assume will be filled with a He:$\rm SF_6$ admixture of 755:5 Torr, with density of 2 $\times 10^{-4}$ $\rm g/cm^3$~\cite{CYGNUS:2020pzb}.
We also assume there will be 100 TPC modules, each of size 10 m$^3$, giving $R_{\rm fid}$ = 6.2 m. 
Each DM-induced nuclear recoil generates a track of ions, which is then drifted in an electric field to the readout plane, where the 2-D (say, $x$-$y$) position of the interaction vertex may be reconstructed.
In addition, using time delays between the arriving pulses along the track, the inclination of the DM particle with respect to the read-out plane may be inferred. 
A typical multiscatter signal would comprise of several such ionization tracks distributed throughout the detector.

Note, however, that this still does not generically reconstruct the absolute $z$ position.
One way to overcome this limitation, discussed in Ref.~\cite{Lewis_2015}, is to use the distribution of drift charge on the readout plane and infer the $z$ position.
This relies solely on the resolution of the readout plane, without requiring timing information.
Another is to use ``minority carriers" such as $\rm SF_5^{-}$, that, due to their drift speeds being different from the main carriers, produce minor peaks near the main peak at readout~\cite{CYGNUS:2020pzb}. 
The time separation between the minor peaks along with the drift speeds can be used to infer the absolute $z$. 
However, the timing resolution of the detector may limit this technique.
A multiscattering DM track may not be fully reconstructed if information on the $z$ positions of the individual recoil sites is missing, unless of course the drift charge distributions just mentioned can help with localizing the $z$ positions.

Our $\sigma_{\chi N}$ vs $m_\chi$ sensitivity requiring a multiplicity of at least 2 has the same log-log slope as DUNE and Xe-1kT, but is orders of magnitude weaker due to the smaller density of the gas in CYGNUS. 
However, the advantage now is that the ceiling on $\sigma_{\chi N}$ could be higher for the same reason.
We obtain the ceiling by requiring that the mean free path of the DM particle in the detector exceeds the 200~$\mu$m resolution of the strip readout~\cite{CYGNUS:2020pzb}:
\begin{align}
\begin{split}
\label{eq:spatialcyg}
\frac{1}{\sigma_{\chi N}^{\rm eff}n_T} \geq 200~{\rm \mu m}~,
\end{split}
\end{align}
where $n_T$ is the number density of the target nuclei.

Coming to DM direct detectors, for XLZD we take a fiducial mass of 30 tonne, implying $R_{\rm fid}$ = 1.3 m, and again assume an LZ-like multi-scatter search. 
That the $\sigma_{\chi N}$ vs $m_\chi$ reach accounts for Poisson fluctuations in multiplicity is a major improvement in the presentation of future sensitivities over the simple horizontal lines shown in Ref.~\cite{Bramante:2018qbc}.
The ceiling on $\sigma_{\chi N}$ here comes from considering  the timing resolution of the detector.
In a dual-phase TPC the interaction vertex is inferred from the time elapsed between the prompt (S1) and proportional (S2) scintillation pulses. 
Above some cross section threshold the S1 pulses begin to merge, so that this timing information is lost.
Using LZ's timing resolution of 200 ns for the S1s to  merge, we obtain the ceiling by requiring
\begin{align}
\begin{split}
\label{eq:event}
\frac{1}{\sigma_{\chi N}^{\rm eff}n_T\bar{v}} \geq 200~{\rm ns}~.
\end{split}
\end{align}
For the liquid argon-based Argo we assume 300 tonne fiducial mass~\cite{ARGO:2018,ARGOSnowmassLOI}, giving $R_{\rm fid}$ = 3.7 m, and otherwise assume that the search will be performed like done at our DUNE module, that is, it will have the signatures and limitations of the DEAP-3600 multiscatter search.

We only show limits for $m_\chi > 10^{16}$~GeV so as to focus on the region unconstrained by former experiments. 
The constraints we do show either subsume or lie far below the constraints in Refs.~\cite{KavanaghSuperheavy:2017cru,gascloudcompendium:Bhoonah:2018gjb,PlasticEtch:Bhoonah2020fys,snowmass:Carney:2022gse,DAMA1999,EdelweissCDMSAlbuquerqueBaudis2003,EdelweissCDMSAlbuquerqueBaudis2003,KavanaghSuperheavy:2017cru,PICO:MS:Broerman2022,XENON1T:MSSSprojexn:Clark:2020mna,CollarBeacomCappiello2021,mica:pricesalamon1986,PlasticEtch:Bhoonah2020fys}.
We also do not show the ceiling to underground searches coming from the Earth overburden as it is far above the maximum cross sections we can reach, at $\sigma_{\chi N}/m_\chi \simeq 10^{-14}{\rm cm^2}/10^{16}~{\rm GeV}$~\cite{KavanaghSuperheavy:2017cru,Bramante:2018qbc,Bramante:2018tos}.
As our reaches lie well below the overburden cross section, in our region of interest the DM kinetic energy would have degraded negligibly after passing through the Earth's crust, since the recoil energy $\sim m_{T}v_\chi^2 \ll$ the kinetic energy $m_\chi v_\chi^2/2$.
This also means that the scattering angle $\sim m_{\rm T}/m_\chi < 10^{-14}$, so our DM would travel in near-straight lines and produce collinear recoils in the detector. 
These may be seen as tracks if the interaction vertices are reconstructed precisely.
Finally, we also do not show the future projections of liquid scintillator-based neutrino experiments such as SNO+ if a DM multiscatter search is undertaken by them, but these can be read from the plots in Ref.~\cite{Bramante:2018tos,Bramante:2019yss}. 

\subsection{Possible reconstruction of velocities}
\label{subsec:velreconstruct}

To perform astrometry of the DM halo as in Ref.~\cite{Bramante:2019yss}, we require reconstruction of the DM velocities event by event.
In Ref.~\cite{Bramante:2019yss} it was pointed out that this was possible at liquid scintillator detectors like SNO+ and JUNO by identifying ``hotspots" in the PMT array that indicate the entry and exit points of the DM particle, and by using the scintillation photons' time-of-flight information to gauge the DM speed.
Here we discuss the possibility of analogous measurements at the detectors considered here.

If the DUNE module were a dual-phase TPC, and if in addition to a PSD-based waveform analysis both S1 and S2 pulses are deployed, then the DM track (more precisely, every recoil juncture in the track) and velocity may be reconstructed as done in, {\em e.g.}, the LZ multiscatter search~\cite{LZ:MS:2024psa}.
The use of ARIADNE cameras (as opposed to SiPMs) would improve the precision of reconstructing multiple sites of recoil~\cite{DUNEModuleDM:PNL2020,*DUNEModuleDM:snowmass:Avasthi2022,*DUNEModuleDM:Bezerra2023}; such a technology would use a camera based on the Timepix3 chip, which would simultaneously measure the position, energy and time-of-arrival of every detected photon, and combine it with an image intensifier~\cite{ARIADNE:Roberts:2018sww}.
Track reconstruction may be done using information on which of the PMTs are fired, and time delays between S1 \& S2 as well as between the S1s arising from multiple scatters.
The same can be said for Xe-1kT and Argo, while  XLZD is already designed to be a dual-phase TPC in the vein of its present-day predecessors.
We do note that reconstructing velocities might be more challenging in liquid argon than in liquid xenon detectors due to slower scintillation response, which is about 2~$\mu$s \cite{PhysRevD.103.043001}, about the same timescale as the maximum interval between multiple nuclear recoils.
For this reason, we also expect the cross section ceiling obtainable from the timing resolution to be lower than the values where there is appreciable multiplicity in scatters.

For CYGNUS, velocity reconstruction might be limited due to the large $\mathcal{O}(1)$~ms drift timescale of charge carriers. 
This is much longer than the DM's detector transit time of about $\mathcal{O}(1) \, \mu$s. 
Therefore timing information of the DM particle traversing the detector might get washed out.
However, we suspect that a resolution of 10~$\mu$s must be within CYGNUS' capability.
This follows from the $\mathcal{O}(1)$~mm length of each ionization track and the 0.14~mm/$\mu$s drift speed of charge carriers~\cite{CYGNUS:2020pzb}, which gives the time delay between pulses as $\mathcal{O}(10) \, \mu$s.
This time delay is already used to reconstruct the inclination of the incident DM, as mentioned before.
Another possibility for
progress here is to use the faster drift of electrons as opposed to heavier charge carriers, which would improve on timing but probably at the cost of losing absolute $z$ reconstruction capability~\cite{LBignell}.
Yet another is to calibrate the detector response to transit times of multiscattering particles like neutrons.
If both the interaction vertices and transit time of the DM track are reconstructed, the velocity vector of the DM is determined.
These speculations require further dedicated study.

\begin{table*}[t]
\setlength{\tabcolsep}{10pt}
\renewcommand{\arraystretch}{1.5}
\begin{tabular}{l l  l  l}
substructure & mean ($v_x$, 
 $v_y$, $v_z$) (km/s) &  \textrm{($\sigma_x$,$\  \sigma_{y}$, $\sigma_z$)  (km/s)} & $\cos^{-1}({\hat{x}_{\text{lab}}\cdot \hat{v}_{\rm str}})$
 \\ \hline \hline
RgDTG-28 & ($-$4, $-$106.1, $-$143.2) & (115.8, 29.3, 30.3)) &  $128.1^{\circ}$\\[1mm]
Sausage & (2.1, $-$0.3, $-$8.7) & (136.6, 35, 72.3) &  $92.8^{\circ}$\\[1mm]
HelmiDTG-1 & (4.5, 197.2, 244.3) & (146, 62.6, 42.4) & $49.5^{\circ}$ \\[1mm]
PgDTG-2 & (221.2, 155.7, 139.7) & (26.2, 33.8, 52.3) &  $56.2^{\circ}$\\[1mm]
SequoiaDTG-4,5 & ($-$36.9, $-$273.9,$-$87.0) & (138.2, 36.7, 65.0)&  $163.32^{\circ}$\\[1mm] \hline
$\vec{v}_{\rm lab}$ & (11.0, 232.3, 7.1) & $-$ &  $-$ \\[1mm] \hline
\end{tabular}
\caption{\label{tab:tablestreams}
Tidal streams considered in this work, taken from Ref.~\cite{streamLAMOST2023}, characterized by the components of their mean and dispersion speeds, and their angle of entry with respect to the mean direction of the lab in the Galactic frame.
Also provided for comparison is the annually averaged lab velocity in the Galactic frame. 
High-speed streams that pass through the Earth at oblique angles are the easiest to discern, as evidenced by the results of our statistical tests.
See Sec.~\ref{subsec:mediantest} for further details.
}
\end{table*}

\section{Tests for streams}
\label{sec:streams}

According to $N$-body and hydrodynamical simulations of galaxy formation, the phase space of the Galactic halo is not smooth and isotropic, but rather contains several unvirialized substructure features. 
These include tidal streams, dark disks, debris flows, and ``shadow bars" trapped by stellar bars~\cite{tidalstreamHelmi:1999ks,tidalstreamFreese:2003tt,darkdiskRead2009,darkdiskMcCullough:2013jma,darkdiskKuhlen:2013tra,darkdiskSchaller:2016uot,myeong2018shardsomegacentauri,shadowbarPetersen2016,shadowbarPetersen:2016vck,dynamicalrelicsYuan2020}; velocity ellipsoid structures from debris flow like the Gaia sausage~\cite{gaiasausage:Necib:2018iwb,gaiasausage:Necib:2018igl,gaiasausage:Buch:2019aiw} have also been observed.
We focus on tidal streams, which arise from tidal stripping of sub-halos by the Milky Way during accretion.
The stellar component of these streams have been inferred in the data of sky surveys such as SDSS, Gaia and LAMOST, often with cross-correlation~\cite{streamSDSSGaia2017,streamGaias2018,streamGaia2020,streamGaiaDR32022,streamLAMOST2023}.
It is possible that a DM component -- even an extended one -- is associated with these streams, and that it may pass through the solar system, notably in the case of the Sagittarius stream~\cite{streamSagittIbata:2000ys,streamSagittBelokurov:2006kc,streamSagittPurcell2012,streamSagittSollima2014}.
While none of these possibilities may be stated with certainty, they may be tested in the laboratory if the DM particles leave multiscatter signals of the nature discussed in this work.

The velocity distributions of streams are usually parameterized as a simple Gaussian added on to the background halo distribution. 
For a single stream that makes up a fraction $\delta$ of the local DM density, we have in the Galactic frame
\begin{eqnarray}
\label{eq:streamdistribution}
f_{\text{tot}}(\vec{v})&=& (1-\delta)f_{\text{halo}}(\vec{v}) + \delta f_{\text{str}}(\vec{v})~,\\
\nonumber f_{\text{str}}(\vec{v}) &\propto& \exp\left( \sum_{i=1}^{3}\frac{(v_i-\mathbf{\mu}_i)^2}{2\sigma_i^2}\right)\Theta(v_{\text{esc}}-|{\vec{v}}|)~,
\end{eqnarray}
where $\mu_i$ and $\sigma_i$ are respectively the $i$th component of mean and dispersion velocities,
and $f_{\rm halo}$ is taken as Eq.~\eqref{eq:velocitydistribution}.
For our analysis we have taken five representative streams from Ref.~\cite{streamLAMOST2023}, summarized in Table~\ref{tab:tablestreams}.

In the following, we perform  a ``non-parametric test" to distinguish the presence of tidal streams over a smooth halo background.
Such a test is applicable to any velocity distribution model for the background halo and substructure components.
We will largely follow the treatment of Ref.~\cite{OHare:2014nxd}.
In that work, the velocity vectors of the incident DM flux are partially reconstructed from the recoil energies and lab-frame scattering angles of the target nucleus, {\em i.e.}, using the techniques of directional detection. 
This in turn is obtainable in a gas TPC from the partial reconstruction of the ionization track induced by a WIMP recoil.
In Ref.~\cite{Kavanagh:2016xfi} this was extended to nuclear recoils in xenon and fluorine, both with and without directional recoil information, and by using an empirical parameterization -- with functional form unknown -- of the DM velocity distribution.
Similarly, in Ref.~\cite{Maity:2022enp} electron recoils were considered, and distributions of recoil energies translated to the number of observable electrons were used to identify streams.
The difference between these approaches and ours is that, as mentioned before, we can directly reconstruct the track and velocity vector of every DM event if sufficient sensitivity to vertex reconstruction and timing is available, as discussed at length in Ref.~\cite{Bramante:2019yss} and in Sec.~\ref{subsec:velreconstruct} here.
In this sense, our statistical test is more directly applicable to the empirically obtained DM velocity distribution.

\begin{figure*}
\includegraphics[width=0.458\textwidth]{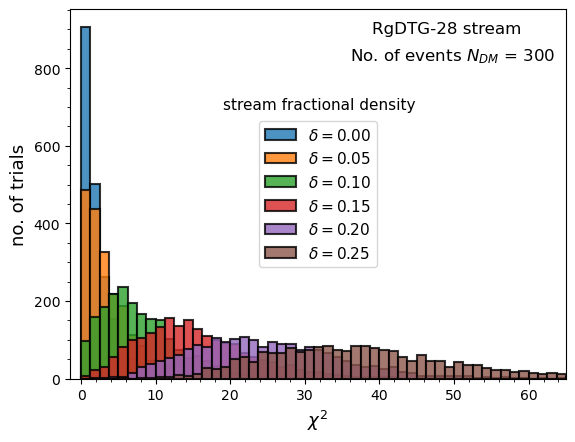} \
\includegraphics[width=0.47\textwidth]{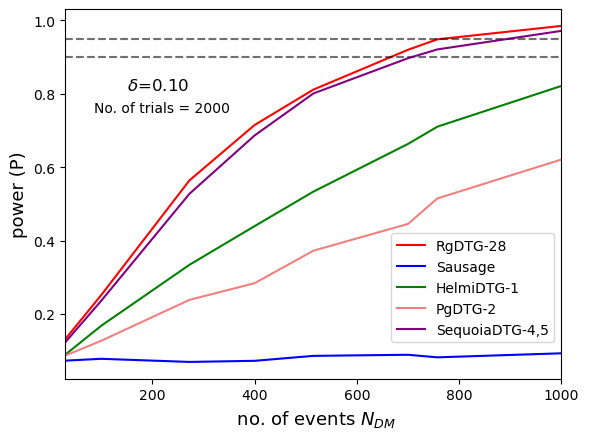}
\caption{\label{fig:rgstreamdist} {\bf \em Left.} The $\chi^2$ distribution (Eq.~\eqref{eq:teststatischisq}) over 2000 pseudodata trials of the median direction hypothesis test performed in Sec.~\ref{subsec:mediantest}, for various fractional DM densities of the RgDTG-28 stream assuming 300 multiscatter events are collected and their velocity vectors reconstructed. 
{\bf \em Right.} The statistical power (Eq.~\eqref{eq:testpower}) of median direction hypothesis tests on various streams, fixing their fractional density to 0.1 and varying the number of multiscatter DM events collected. 
For reference we have marked $P$ = 0.90 and $P$ = 0.95 with horizontal dashed lines.
Streams that are faster relative to the Earth and incident more obliquely are picked out better. 
See Sec.~\ref{subsec:mediantest} for further discussion.
}
\end{figure*}

\subsection{Median direction hypothesis test}
\label{subsec:mediantest}

We now perform the median direction hypothesis test discussed in Refs.~\cite{1989Ap&SS.151..177M,OHare:2014nxd}.
For a set of $N$ Galactic frame unit velocity vectors  $\hat{x}_i$ in a data sample, the median direction $\hat{x}_{\text{med}}$ is defined as that which minimizes
\begin{align}
\begin{split}
\label{eq:minimizemedian}
\sum_{i=1}^{N}\,\cos^{-1}({\hat{x_i}\cdot \hat{x}_{\text{med}}})~.
\end{split}
\end{align}
The median direction hypothesis test then checks if the sample median $\hat{x}_{\rm med}$ is consistent with a hypothesis median $\hat{x}_0$.
Note that the {\em magnitude} of the velocity need not be reconstructed for this test, {\em i.e.} the timing resolution of the detector is not essential. 
So long as the directional information is obtained as discussed in Secs.~\ref{subsec:reachesxsvmx} and \ref{subsec:velreconstruct}, this test can be performed.

For a smooth halo, we expect that the median direction of the sampled events will be opposite to the lab's velocity, $-\hat{x}_{\text{lab}}$.
In general this velocity is time-dependent,
but since the average galactocentric speed of the Earth, 
30 km/s~\cite{Baxter:2021pqo}, is much smaller than the solar motion, $|\vec{v}_{0}+\vec{v}^\odot_{\rm pec}|\approx$ 252 km/s, the direction of the lab velocity only varies by small angles. 
Thus to a good approximation the median direction is the one opposite to solar motion in the Galactic frame.
In practice we take $- \hat{x}_{\text{lab}}$ averaged over a year as the hypothesized median. 

For polar coordinates of the sample median direction ($\theta_{\text{med}},\phi_{\text{med}})$,
we first transform the co-ordinates using rotation matrices around the $y$ and $z$ axes,
\begin{align}
\begin{split}
\label{eq:rotationmedianhyp}
\hat{x}_i' = R_y(\theta_{\text{med}})R_z(-\phi_{\text{med}})\hat{x}_i~,
\end{split}
\end{align}
and then using azimuthal angles $\phi_i'$ 
in the new coordinates construct the matrix
\begin{eqnarray}
\label{eq:medianmatrix}
&& \Sigma=\frac{1}{2}
\begin{pmatrix}
    \sigma_{11} & \sigma_{12}\\
    \sigma_{21} & \sigma_{22}
\end{pmatrix}~,\\
\nonumber \sigma_{11}&=&1+\frac{1}{N}\sum_{i=1}^{N}\cos2\phi_i'~, \ \
\sigma_{22}=1-\frac{1}{N}\sum_{i=1}^{N}\cos2\phi_i'~,\\
\nonumber \sigma_{12}&=&\sigma_{21}=\frac{1}{N}\sum_{i=1}^{N}\sin\phi_i'~.
\end{eqnarray}
Next, we again rotate the vectors $\hat{x}_i$ using Eq.~\eqref{eq:rotationmedianhyp}, but now measured relative to a hypothesized median direction ($\theta_{0}$,$\phi_{0}$).
The azimuthal angles in the new co-ordinates are now  $\phi_i''$. 
We then construct the test statistic as
\begin{eqnarray}
\label{eq:teststatischisq}
\nonumber \chi^2=U^T\Sigma^{-1}U~, \\
U=\frac{1}{\sqrt{N}}
\begin{pmatrix}
    \sum\cos\phi_i''\\
    \sum\sin\phi_i''~
\end{pmatrix}~.
\end{eqnarray}

To demonstrate this test we take the stream RgDTG-28 
and generate 2000 sets of \{$x_i$\} sampled from the distribution in Eq.~\eqref{eq:streamdistribution} after weighting it appropriately with the DM velocity: see the appendix for a description of this procedure. Here every sampled (time-dependent) velocity vector is transformed to the Galactic frame. 
In the left hand panel of Fig.~\ref{fig:rgstreamdist} we show the distribution of $\chi^2$ across the 2000 trials,
fixing the number of events $N_{\rm DM} = 300$ while varying the stream fractional density $\delta$.
The null hypothesis corresponding to $\delta = 0$ is seen to peak at $\chi^2 = 0$, whereas as $\delta$ is increased the peak shifts to higher $\chi^2$ while the distribution itself broadens. 
Such a broad peak is obtained -- as opposed to a narrow one --  because we are testing the presence of a stream as an {\em addition} to the smooth background halo (Eq.~\eqref{eq:streamdistribution}), so there is always a non-negligible probability that the measurement mimics a standard halo scenario.

We can now estimate the {\em statistical power} of the median direction hypothesis test.
Our null hypothesis here, $p_0$, is a smooth halo, and the alternative hypothesis $p_1$ is a halo + stream, where $p_0$ and $p_1$ are probability distributions obtained over many pseudodata trials, as done in the median direction hypothesis test. 
In general, for a test statistic $\lambda$, the ``test size" $\alpha$ is
\begin{align}
\begin{split}
\label{eq:testsizestat}
\alpha=\int_{\lambda_{\rm crit}}^{\infty}p_0(\lambda)\,d\lambda~,
\end{split}
\end{align}
which is the probability of measuring $\lambda > \lambda_{\rm crit}$ if the null hypothesis is correct. 
Requiring 95\% confidence level, we set $\alpha=0.05$, which then determines $\lambda_{\rm crit}$. 
For our median direction hypothesis test, the test statistic $\lambda$ is nothing but the $\chi^2$ in Eq.~\eqref{eq:teststatischisq}.
The statistical power is then defined as the probability of rejecting the null hypothesis if it is false. 
Said differently, it is the fraction of an ensemble of experiments in which a stream can be detected with $\lambda > \lambda_{\rm crit}$ if the alternative hypothesis is true.
It is given by
\begin{align}
\begin{split}
\label{eq:testpower}
P=\int_{\lambda_{\rm crit}}^{\infty}p_1(\lambda)\,d\lambda~.
\end{split}
\end{align}

In the right panel of Fig.~\ref{fig:rgstreamdist} we show the power obtained by varying the number of observed multiscatter events while fixing the stream fractional density $\delta = 0.1$ for our benchmark streams in Table~\ref{tab:tablestreams}.
Fewer events are required to pick out streams that have both a high relative speed $= |\vec{v}_{\rm str} - \vec{v}_{\rm lab}|$ and an oblique angle of incidence at the laboratory, $\cos^{-1}({\hat{x}_{\text{lab}}\cdot \hat{v}_{\rm str}})$.
Faster streams would simply source more events due to their contribution to the overall velocity distribution in Eq.~\eqref{eq:streamdistribution}, and oblique streams have a higher chance of being picked up in the median direction hypothesis test.  Thus RgDTG-28 and SequoiaDTG-4,5 are the easiest to discern, followed by HelmiDTG-1 and PgDTG-2, whereas the very slow Sausage stream is quite challenging to detect.

To sum up, both the left and right panels of Fig.~\ref{fig:rgstreamdist} illustrate the fact that statistical testing of the presence of streams improves with an increase in both their fractional density $\delta$ and the number of multi-scatter events collected.

\section{Discussion}
\label{sec:concs}

In this study we worked out the sensitivities of future large-volume detectors such as the far-detector ``module of opportunity" of DUNE, 
kilotonne liquid xenon detectors that are proposed for $0\nu\beta\beta$ decay searches,
the proposed gaseous directional recoil detector CYGNUS, and the dark matter detectors XLZD and Argo.
We expect the reach of PandaX-xT~\cite{PandaX-xT:2024oxq} to be similar to that of XLZD.
We also showed that, in cases where DM velocity vectors may be reconstructed, not only can broad properties of the halo velocity distribution be marked, but the presence of halo substructure such as tidal streams may be inferred with suitable statistical tests. 
Our findings thus significantly update the program of dark matter direct detection in the high-mass, multiscatter regime, which was previously restricted to dark matter detectors and liquid scintillator-based neutrino detectors.
While we adapted the CYGNUS design to illustrate multiscattering DM sensitivities of gaseous TPC detectors, one could also consider 10$-$100 meter scale detectors that use large volumes of ultra-high pressure gas stored in underground caverns~\cite{SnowmassNuFog:2022ort}.

Our work complements Ref.~\cite{Acevedo:2024wmx}, which also considers future large-volume detectors including DUNE, but seeking signatures of MeV$-$GeV scale DM boosted by an attractive fifth force with the Earth. 
Such studies and ours underline the importance of repurposing proposed experimental designs in the service of discovering and studying particle dark matter.

\section*{Acknowledgments}

We thank Ranjan Laha for a talk at WHEPP XVII that inspired this study.
We also gratefully acknowledge Ciaran O' Hare and Lindsey Bignell for multiple helpful inputs on CYGNUS, and 
Nahuel Ferreiro Iachellini,
Chris Kouvaris,
Tarak Nath Maity,
Ibles Olcina Samblas,
and
Ryan Smith
for discussion. 
\\

\appendix*

\section{\label{sec:fluxestimate}Estimate of the dark matter flux in non-spherical fiducial volumes}

The integrated flux in Eq.~\eqref{eq:fluxsphere} that we use to derive our scattering cross section vs mass reaches in Fig.~\ref{fig:xsvmreaches} assumes a fully spherical geometry for the detector.
It is this assumption that in Eq.~\eqref{eq:fluxsphere} gives the factor of $\pi R_{\rm fid}^2$, the geometric cross section of a sphere. 
While spherical detectors present a simple case, it might be more challenging to obtain the DM flux for other geometries.
In this appendix we show one way to treat non-spherical geometries.
The main subtlety is the interdependence of the geometry and the anisotropy of the velocity distribution in the lab frame.
In the following we will use the notation $\tilde{f}$ to denote flux-weighted speed distributions, $f$ to denote velocity distributions, and $\bar{f}$ to denote speed distributions after the angles are integrated out. 
We will also use $\Phi$ to denote the integrated flux that is obtained after integrating over the speeds in $\tilde{f}$ such as in Eq.~\eqref{eq:fluxsphere}.

\subsection{Basic setup}

Let us begin with a simple picture for estimating the DM flux.
Consider DM with {\em uniform} velocity $\vec{v}_\chi$ so that the velocity distribution is 
\begin{equation}
 f_{\rm unif}(\vec{u}) = \delta^3 (\vec{u} - \vec{v}_\chi)~.
 \label{}
\end{equation}
Then for DM density $\rho_{\chi}$ the integrated flux over an exposure time $t_{\rm exp}$ through a plane surface with area $A_{\rm det}$ perpendicular to the velocity field is given by
\begin{eqnarray}
\nonumber \Phi_{\rm unif} &=& (\rho_{\chi}/m_\chi)A_{\text{det}}t_{\text{exp}}~\int d^3 u \,u f_{\rm unif}(\vec{u}) \\
&=& (\rho_\chi/m_\chi)A_{\text{det}} v_\chi t_{\text{exp}}~. 
\label{eq:intfluxvfield}
\end{eqnarray}
Realistically, we must integrate velocities over a distribution, and account for the angles at which the velocity vectors
intersect the detector surface. 
For a general detector surface, we therefore have
\begin{eqnarray}
\label{eq:gensurfaceflux}
\Phi &=& (\rho_{\chi}/m_\chi)t_{\text{exp}} \times \\
\nonumber && \int dA_{\rm det} \int_{\mbox{\tiny{0}}}^{\mbox{\tiny{$\infty$}}}du\! \int_{\rm in} d\Omega \ u (\hat{u} \cdot \hat{n})u^{2}f(\vec{u}),~
\end{eqnarray}
where the first integral is over the detector surface,
the ``in" in the third integral denotes integrating over only the velocity solid angles that provide flux {\em into} the detector, and $\hat{n}$ is the local normal to the surface pointing inward.

Now if in Eq.~\eqref{eq:gensurfaceflux} we had neglected the $\hat{u} \cdot \hat{n}$ term, and evaluated the integrals over $A_{\rm det}$ and $\Omega$ independent of each other, we would have obtained the integrated flux as
\begin{equation}
    \Phi_{\rm ave} = (\rho_\chi/m_\chi) S_{\text{det}} \bar{v}_\chi t_{\text{exp}}~,
    \label{eq:integfluxavespeed}
\end{equation}
where $\bar{v}_\chi$ is the average speed = $\int d\Omega \ du \ u^3 f(\vec{u})$, and $S_{\rm det}$ is the total surface area of the detector.
Note that the form of this integrated flux looks similar to Eq.~\eqref{eq:gensurfaceflux} with the replacements $v_\chi \to \bar{v}_\chi$ and $A_{\rm det} \to S_{\rm det}$.
For a spherical detector this would overestimate the integrated flux by a factor of 4, since $S_{\rm det} = 4\pi R^2_{\rm det}$, while one must instead use the geometric cross sectional area $\pi R_{\rm det}^2$.
This results from the $\hat{u} \cdot \hat{n}$ factors, which ensure that for every $\hat{u}$ the effective entry area on the spherical detector is $\pi R^2_{\rm det}$.

In the next subsection we will see how to obtain the integrated flux for detector fiducial volumes that are {\em nearly} spherical.
Following this, we will also treat cuboidal detectors.
While the integrated flux in Eq.~\eqref{eq:gensurfaceflux} 
can be estimated by brute force for any detector geometry,
in practice we estimate it by identifying relations between $\hat{u} \cdot \hat{n}$ and the location on the detector.

We now remark on a subtlety regarding reconstructing the DM velocity distribution empirically in multiscatter searches.
In Ref.~\cite{Bramante:2019yss} events were sampled from a flux-weighted speed distribution $\tilde{f}$, before they were unweighted by a factor of the speed $u$ and the resulting speed distribution $\bar{f}$ was used for identifying the mean speed, escape speed, and so on.
This was possible because detectors were assumed to be fully spherical, which simplifies the treatment of the $\hat{u} \cdot \hat{n}$ in Eq.~\eqref{eq:gensurfaceflux} as just discussed.
For non-spherical detector geometries, however, the $\hat{u} \cdot \hat{n}$ factor would make the procedure more non-trivial.
We cannot easily unweight the flux-weighted distribution -- as obtained at an experiment -- to get the actual halo speed distribution. 
In principle, we could try to reconstruct the velocity distribution as opposed to the speed distribution with angles integrated out, but this would mean getting a 2-D distribution over ($u$, $\theta$), degrading the statistical precision of the analysis. 
Having said all this, we find in practice that 
taking the flux-weighted distribution as that of a spherical detector
is still a very good approximation -- even for a cuboidal detector -- at the few percent level, and hence the procedure outlined in Ref.~\cite{Bramante:2019yss} may be carried out to broadly mark the properties of the DM velocity distribution.

\subsection{Near-spherical detectors}
\label{subsec:spherdetflux}

In practice, the volume of detector liquid used for performing a multiscatter DM search need not be perfectly spherical.
This may be because the fill level may not be all the way to the top of a spherical vessel, as in DEAP-3600~\cite{DEAP:SS:2019yzn,DEAP:MS:2021raj}.
One could expect something similar at SNO+~\cite{SNOPlus:2021xpa} and JUNO~\cite{JUNO:2021vlw} as well.
Of course, the vessel itself may not be spherical, and fiducial volumes chosen may naturally take on non-spherical shapes, such as in multiscatter searches at XENON1T~\cite{XENON1T:MS:2023iku} and LZ~\cite{LZ:MS:2024psa}.
In this sub-section we provide a prescription for evaluating the integrated flux of DM in these geometries.

The infinitesimal flux at some point on the detector surface multiplied by the area element $dA$ is given by
\begin{align}
\begin{split}
\label{eq:isotropicsphere}
d\tilde{f}_{\rm sph}=(u\cos{\theta})u^{2}f (u, \theta, \phi) d\cos\theta \,d\phi\, du \, dA~,
\end{split}
\end{align}
where the factor of $u\cos\theta$ plays the role of $\hat u \cdot \hat n$ in Eq.~\eqref{eq:gensurfaceflux}.
For a velocity distribution $f (u, \theta, \phi)$ that is isotropic, the integral over $dA$ factorizes and gives $4\pi R_{\text{det}}^2$ for a spherical detector radius $R_{\rm det}$.
This is a simplification used previously in computing the DM flux through stellar bodies; see, {\em e.g.}, Refs.~\cite{Kouvaris_2008,Bramante_2017}.
For a non-spherical fiducial volume of surface area $S_{\rm fid}$, one can get the integrated flux by replacing $\int dA \to S_{\rm fid}$.
However, for an anisotropic $f (u, \theta, \phi)$, such as the Maxwell-Boltzmann distribution boosted to the Earth's frame, we must be more careful, as different points in the detector may now receive different amounts of flux.

In addition to $\theta$ and $\phi$ that describe polar and azimuthal angles in {\em velocities}, we introduce the angles $\theta_\omega$ and $\phi_{\omega}$ to describe {\em positions} on the detector in spherical polar co-ordinates, taking the center of the detector as the origin.
For simplicity, we choose the polar co-ordinates of $(\theta_\omega,\phi_\omega)$ to align with those of $(\theta,\phi)$.
We now have the flux-weighted speed distribution as 
\begin{eqnarray}
\label{eq:spherefluxdist}
\nonumber \tilde{f}_{\rm \sim sph}(u) &=&  \int \int r^2(\theta_\omega, \phi_\omega) d\phi_{\omega} d \cos{\theta_\omega}\ \\ && 
\left(\int_{\rm in} d\Omega \  u^{3}|\cos{\theta_{\text{\tiny{loc}}}}|f(u, \theta, \phi)\right)~,
\end{eqnarray}
where $|\cos{\theta_{\text{loc}}}|$ depends locally on the angle between the incoming DM direction and the normal at the detector surface, {\em i.e.} it is the $\hat{u}\cdot\hat{n}$ in Eq.~\eqref{eq:gensurfaceflux}, with the absolute value ensuring that we only account for velocity vectors pointing inward.
The key difference between Eqs.~\eqref{eq:isotropicsphere} and Eq.~\eqref{eq:spherefluxdist} is that the $(\theta,\phi)$ integral in the latter depends on the location on the sphere $(\theta_\omega,\phi_{\omega})$.

\begin{figure}
\includegraphics[width=0.47\textwidth]{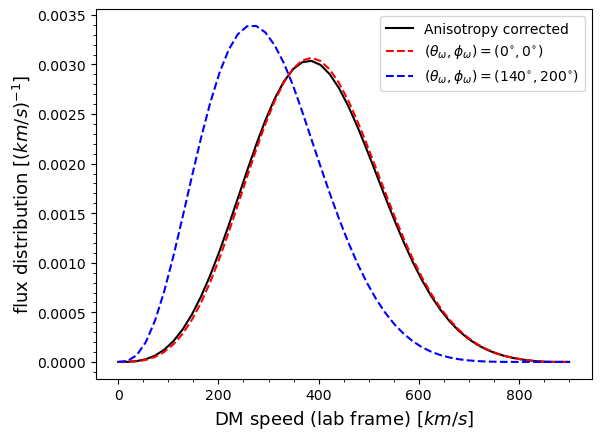}  \
\\
\caption{\label{fig:spheredist}  
Normalized flux-weighted speed distributions for spherical detectors, illustrating discrepancies that may arise from not accounting for interdependences between anisotropies in the velocity distribution and the location of the detector at which dark matter is incident. 
The solid black curve is obtained from our updated treatment discussed below Eq.~\eqref{eq:spherefluxdist}, and the dashed red and blue curves from Eq.~\eqref{eq:isotropicsphere} for the various detector locations indicated.}
\end{figure}

Before working out this interdependence of $\theta_{\rm loc}$, $\theta_\omega$ and $\phi_\omega$, we first illustrate its importance in Fig.~\ref{fig:spheredist}.
A fully spherical fiducial volume is assumed here for simplicity.
The solid black curve is the normalized distribution after performing all the angular integrals in Eq.~\eqref{eq:spherefluxdist} for a Maxwell-Boltzmann $f$ that is boosted to the lab frame.
The dotted curves show the normalized lab-frame distribution in Eq.~\eqref{eq:isotropicsphere}, with blue corresponding to the detector location $(\theta_\omega, \phi_\omega)$ = $(140^{\circ},\,200^{\circ})$ and red to $(0^{\circ},\, 0^{\circ})$. 
The maximum fractional discrepancy between the former and the correct distribution (Eq.~\eqref{eq:spherefluxdist}, black curve) is 157\%, while for the latter it is only 6\%.
This underlines the wide range of errors that may arise from the use of an incorrect treatment of flux.

Let us now take as a concrete example a spherical detector filled up to some level with some liquid, {\em e.g.} LAr or organic scintillator. 
The flux through the flat surface on the top is estimated easily by identifying the correct limits of integration of $\theta_\omega$ and $\phi_\omega$ in Eq.~\eqref{fig:spheredist}, and including only the lower hemisphere of velocity vectors weighted by $f$.
To estimate the net flux through the rest of the curved surface, we must identify a relation between $\cos\theta_{\rm loc}$,  ($\theta, \phi$) and ($\theta_\omega, \phi_\omega$).
One way to do this is to notice that the relevant velocity vectors, which result in an inward flux, comprise a hemisphere at every point on the detector and thereby identify the integration limits on ($\theta$, $\phi$) in terms of $(\theta_\omega, \phi_\omega)$.
However, in practice obtaining such limits is challenging
nor does this give a clear relation between $\theta_{\rm loc}$ and the other variables. 

We find that a simpler approach is to transform the velocity vectors such that the local normal to the relevant point in the detector is mapped on to the normal at the top of the spherical detector (which, we remind the reader, is not the top of the liquid surface).
In this process all velocity vectors are suitably rotated, and the evaluation of angular integrals becomes straightforward.
These rotations are given by
\begin{align}
\begin{split}
\label{eq:rotspherical}
\vec{v}' = R_z(\phi_{\omega})R_y(-\theta_{\omega})R_z(-\phi_{\omega})\vec{v}~,
\end{split}
\end{align}
where $R_i(\alpha)$ are matrices effecting rotations around the direction $i$ by an angle $\alpha$.
Note here that the co-ordinate system of the velocities is itself not changed, which helps us preserve the form of the distribution $f$.
Once this is done, we can simply set $\theta_{\rm loc} = \theta$, and take the limits on $\theta$ as [$\pi/2,\pi$] and on $\phi$ as [$0,2\pi$].
To explain this in more detail:
for the point ($\theta_\omega, \phi_\omega) = (0,0)$ on the detector, the incoming flux directions lie in the region $\theta \in [\pi/2,\pi]$ and $\phi \in [0,2\pi]$, and crucially, $\theta_{\rm loc} = \theta$.
The transformation in Eq.~\eqref{eq:rotspherical} now maps the normal at this reference point to the normal at any  ($\theta_{\omega}$,$\phi_{\omega}$).
Thus, it establishes a one-to-one mapping between the reference set of velocity vectors and those at another point on the spherical detector. 
Moreover, since the dot product is conserved in rotations, this procedure guarantees that $\theta_{\rm loc}$ is unchanged, that is, we still integrate over the component of the velocity vector along the local normal on the detector. 

 We remark that our prescription may be useful not only for obtaining the integrated flux, but also for reconstructing the angular distribution of DM velocities if multiscatter tracks can be identified, and reveal information on possible anisotropies inherent in the velocity distribution.
 This is again because events registered at a detector are distributed according to the flux-weighted $\tilde{f}$, which contains the detector position-dependent factor of $|\cos \theta_{\rm loc}|$.
 Incidentally, for the median direction hypothesis test for streams described in Sec.~\ref{subsec:mediantest} it is just the flux-weighting seen in Eq.~\eqref{eq:spherefluxdist} that we used for sampling events.

\subsection{Cuboidal detectors}

Examples of large-volume cuboidal detectors are DUNE and CYGNUS.
Labelling the detector faces $\pm \hat{N}$, $\pm \hat{W}$ and $\pm \hat{Z}$ (for ``north", ``west", and ``zenith"), 
we can use the surface flux in Eq.~\ref{eq:gensurfaceflux} directly to obtain the net flux through the detector.
Without loss of generality, we can place the cuboid in spherical polar co-ordinates of the velocity vector
and integrate over appropriate limits on $\theta$ and $\phi$.
 The flux-weighted speed distribution at the detector is then
\begin{align}
\begin{split}
\label{eq:fluxcubical}
\tilde{f}_{\rm cub}(u)= \sum_{i=1}^{6}\,A_{i}\iint_{[i]} d \cos\theta\, d\phi \,(\hat{u}\cdot\hat{n}_i)u^{3}f(u,\theta,\phi)~,  
\end{split}
\end{align}
where [$i$], $A_i$, $\hat{n}_i$ are respectively the integration limits of, area of, and normal to the $i$th face. 
We take the integration limits [$i$] of $(\theta, \phi)$ for each face as the following: 
$\hat{N}: ([0,\pi],[\pi,2\pi])$,
$-\hat{N}: ([0,\pi],[0,\pi])$,
$\hat{W}: ([0,\pi],[\pi/2,3\pi/2])$,
$-\hat{W}: ([0,\pi],[3\pi/2,\pi/2])$,
$\hat{Z}: ([\pi/2,\pi],[0,2\pi])$,
$-\hat{Z}: ([0,\pi/2],[0,2\pi])$.
Other orientations of the detector may be treated with a co-ordinate transformation.
We find that the annual average of the distribution $\tilde{f}_{\rm cub}(u)$ is nearly coincident with the one on March 7th.

\bibliography{refs}

\begin{thebibliography}{90}%
\makeatletter
\providecommand \@ifxundefined [1]{%
 \@ifx{#1\undefined}
}%
\providecommand \@ifnum [1]{%
 \ifnum #1\expandafter \@firstoftwo
 \else \expandafter \@secondoftwo
 \fi
}%
\providecommand \@ifx [1]{%
 \ifx #1\expandafter \@firstoftwo
 \else \expandafter \@secondoftwo
 \fi
}%
\providecommand \natexlab [1]{#1}%
\providecommand \enquote  [1]{``#1''}%
\providecommand \bibnamefont  [1]{#1}%
\providecommand \bibfnamefont [1]{#1}%
\providecommand \citenamefont [1]{#1}%
\providecommand \href@noop [0]{\@secondoftwo}%
\providecommand \href [0]{\begingroup \@sanitize@url \@href}%
\providecommand \@href[1]{\@@startlink{#1}\@@href}%
\providecommand \@@href[1]{\endgroup#1\@@endlink}%
\providecommand \@sanitize@url [0]{\catcode `\\12\catcode `\$12\catcode `\&12\catcode `\#12\catcode `\^12\catcode `\_12\catcode `\%12\relax}%
\providecommand \@@startlink[1]{}%
\providecommand \@@endlink[0]{}%
\providecommand \url  [0]{\begingroup\@sanitize@url \@url }%
\providecommand \@url [1]{\endgroup\@href {#1}{\urlprefix }}%
\providecommand \urlprefix  [0]{URL }%
\providecommand \Eprint [0]{\href }%
\providecommand \doibase [0]{http://dx.doi.org/}%
\providecommand \selectlanguage [0]{\@gobble}%
\providecommand \bibinfo  [0]{\@secondoftwo}%
\providecommand \bibfield  [0]{\@secondoftwo}%
\providecommand \translation [1]{[#1]}%
\providecommand \BibitemOpen [0]{}%
\providecommand \bibitemStop [0]{}%
\providecommand \bibitemNoStop [0]{.\EOS\space}%
\providecommand \EOS [0]{\spacefactor3000\relax}%
\providecommand \BibitemShut  [1]{\csname bibitem#1\endcsname}%
\let\auto@bib@innerbib\@empty
\bibitem [{\citenamefont {Goodman}\ and\ \citenamefont {Witten}(1985)}]{GoodmanWitten:1984dc}%
  \BibitemOpen
  \bibfield  {author} {\bibinfo {author} {\bibfnamefont {M.~W.}\ \bibnamefont {Goodman}}\ and\ \bibinfo {author} {\bibfnamefont {E.}~\bibnamefont {Witten}},\ }\href {\doibase 10.1103/PhysRevD.31.3059} {\bibfield  {journal} {\bibinfo  {journal} {Phys. Rev. D}\ }\textbf {\bibinfo {volume} {31}},\ \bibinfo {pages} {3059} (\bibinfo {year} {1985})}\BibitemShut {NoStop}%
\bibitem [{\citenamefont {Carney}\ \emph {et~al.}(2023)\citenamefont {Carney} \emph {et~al.}}]{snowmass:Carney:2022gse}%
  \BibitemOpen
  \bibfield  {author} {\bibinfo {author} {\bibfnamefont {D.}~\bibnamefont {Carney}} \emph {et~al.},\ }\href {\doibase 10.21468/SciPostPhysCore.6.4.075} {\bibfield  {journal} {\bibinfo  {journal} {SciPost Phys. Core}\ }\textbf {\bibinfo {volume} {6}},\ \bibinfo {pages} {075} (\bibinfo {year} {2023})},\ \Eprint {http://arxiv.org/abs/2203.06508} {arXiv:2203.06508 [hep-ph]} \BibitemShut {NoStop}%
\bibitem [{\citenamefont {Chung}\ \emph {et~al.}(1998)\citenamefont {Chung}, \citenamefont {Kolb},\ and\ \citenamefont {Riotto}}]{Models:Chung1998wimpzillas}%
  \BibitemOpen
  \bibfield  {author} {\bibinfo {author} {\bibfnamefont {D.~J.~H.}\ \bibnamefont {Chung}}, \bibinfo {author} {\bibfnamefont {E.~W.}\ \bibnamefont {Kolb}}, \ and\ \bibinfo {author} {\bibfnamefont {A.}~\bibnamefont {Riotto}},\ }\href {\doibase 10.1103/physrevd.59.023501} {\bibfield  {journal} {\bibinfo  {journal} {Physical Review D}\ }\textbf {\bibinfo {volume} {59}} (\bibinfo {year} {1998}),\ 10.1103/physrevd.59.023501}\BibitemShut {NoStop}%
\bibitem [{\citenamefont {Kolb}\ \emph {et~al.}(1998)\citenamefont {Kolb}, \citenamefont {Chung},\ and\ \citenamefont {Riotto}}]{Models:kolb1998wimpzillas}%
  \BibitemOpen
  \bibfield  {author} {\bibinfo {author} {\bibfnamefont {E.~W.}\ \bibnamefont {Kolb}}, \bibinfo {author} {\bibfnamefont {D.~J.~H.}\ \bibnamefont {Chung}}, \ and\ \bibinfo {author} {\bibfnamefont {A.}~\bibnamefont {Riotto}},\ }\href@noop {} {\enquote {\bibinfo {title} {Wimpzillas!}}\ } (\bibinfo {year} {1998}),\ \Eprint {http://arxiv.org/abs/hep-ph/9810361} {arXiv:hep-ph/9810361 [hep-ph]} \BibitemShut {NoStop}%
\bibitem [{\citenamefont {Harigaya}\ \emph {et~al.}(2016)\citenamefont {Harigaya}, \citenamefont {Lin},\ and\ \citenamefont {Lou}}]{Models:Harigaya2016:GUTzillas}%
  \BibitemOpen
  \bibfield  {author} {\bibinfo {author} {\bibfnamefont {K.}~\bibnamefont {Harigaya}}, \bibinfo {author} {\bibfnamefont {T.}~\bibnamefont {Lin}}, \ and\ \bibinfo {author} {\bibfnamefont {H.~K.}\ \bibnamefont {Lou}},\ }\href {\doibase 10.1007/JHEP09(2016)014} {\bibfield  {journal} {\bibinfo  {journal} {JHEP}\ }\textbf {\bibinfo {volume} {09}},\ \bibinfo {pages} {014} (\bibinfo {year} {2016})},\ \Eprint {http://arxiv.org/abs/1606.00923} {arXiv:1606.00923 [hep-ph]} \BibitemShut {NoStop}%
\bibitem [{\citenamefont {De~Luca}\ \emph {et~al.}(2018)\citenamefont {De~Luca}, \citenamefont {Mitridate}, \citenamefont {Redi}, \citenamefont {Smirnov},\ and\ \citenamefont {Strumia}}]{Models:ColoredDM}%
  \BibitemOpen
  \bibfield  {author} {\bibinfo {author} {\bibfnamefont {V.}~\bibnamefont {De~Luca}}, \bibinfo {author} {\bibfnamefont {A.}~\bibnamefont {Mitridate}}, \bibinfo {author} {\bibfnamefont {M.}~\bibnamefont {Redi}}, \bibinfo {author} {\bibfnamefont {J.}~\bibnamefont {Smirnov}}, \ and\ \bibinfo {author} {\bibfnamefont {A.}~\bibnamefont {Strumia}},\ }\href {\doibase 10.1103/PhysRevD.97.115024} {\bibfield  {journal} {\bibinfo  {journal} {Phys. Rev.}\ }\textbf {\bibinfo {volume} {D97}},\ \bibinfo {pages} {115024} (\bibinfo {year} {2018})},\ \Eprint {http://arxiv.org/abs/1801.01135} {arXiv:1801.01135 [hep-ph]} \BibitemShut {NoStop}%
\bibitem [{\citenamefont {Davoudiasl}\ and\ \citenamefont {Mohlabeng}(2018)}]{Models:BNL:DarkBaryonGeVMediator}%
  \BibitemOpen
  \bibfield  {author} {\bibinfo {author} {\bibfnamefont {H.}~\bibnamefont {Davoudiasl}}\ and\ \bibinfo {author} {\bibfnamefont {G.}~\bibnamefont {Mohlabeng}},\ }\href {\doibase 10.1103/PhysRevD.98.115035} {\bibfield  {journal} {\bibinfo  {journal} {Phys. Rev.}\ }\textbf {\bibinfo {volume} {D98}},\ \bibinfo {pages} {115035} (\bibinfo {year} {2018})},\ \Eprint {http://arxiv.org/abs/1809.07768} {arXiv:1809.07768 [hep-ph]} \BibitemShut {NoStop}%
\bibitem [{\citenamefont {Hardy}\ \emph {et~al.}(2015{\natexlab{a}})\citenamefont {Hardy}, \citenamefont {Lasenby}, \citenamefont {March-Russell},\ and\ \citenamefont {West}}]{Models:nucleiHardy:2014mqa}%
  \BibitemOpen
  \bibfield  {author} {\bibinfo {author} {\bibfnamefont {E.}~\bibnamefont {Hardy}}, \bibinfo {author} {\bibfnamefont {R.}~\bibnamefont {Lasenby}}, \bibinfo {author} {\bibfnamefont {J.}~\bibnamefont {March-Russell}}, \ and\ \bibinfo {author} {\bibfnamefont {S.~M.}\ \bibnamefont {West}},\ }\href {\doibase 10.1007/JHEP06(2015)011} {\bibfield  {journal} {\bibinfo  {journal} {JHEP}\ }\textbf {\bibinfo {volume} {06}},\ \bibinfo {pages} {011} (\bibinfo {year} {2015}{\natexlab{a}})},\ \Eprint {http://arxiv.org/abs/1411.3739} {arXiv:1411.3739 [hep-ph]} \BibitemShut {NoStop}%
\bibitem [{\citenamefont {Hardy}\ \emph {et~al.}(2015{\natexlab{b}})\citenamefont {Hardy}, \citenamefont {Lasenby}, \citenamefont {March-Russell},\ and\ \citenamefont {West}}]{Models:nucleiHardy:2015boa}%
  \BibitemOpen
  \bibfield  {author} {\bibinfo {author} {\bibfnamefont {E.}~\bibnamefont {Hardy}}, \bibinfo {author} {\bibfnamefont {R.}~\bibnamefont {Lasenby}}, \bibinfo {author} {\bibfnamefont {J.}~\bibnamefont {March-Russell}}, \ and\ \bibinfo {author} {\bibfnamefont {S.~M.}\ \bibnamefont {West}},\ }\href {\doibase 10.1007/JHEP07(2015)133} {\bibfield  {journal} {\bibinfo  {journal} {JHEP}\ }\textbf {\bibinfo {volume} {07}},\ \bibinfo {pages} {133} (\bibinfo {year} {2015}{\natexlab{b}})},\ \Eprint {http://arxiv.org/abs/1504.05419} {arXiv:1504.05419 [hep-ph]} \BibitemShut {NoStop}%
\bibitem [{\citenamefont {Butcher}\ \emph {et~al.}(2017)\citenamefont {Butcher}, \citenamefont {Kirk}, \citenamefont {Monroe},\ and\ \citenamefont {West}}]{Models:nucleiMonroe:2016hic}%
  \BibitemOpen
  \bibfield  {author} {\bibinfo {author} {\bibfnamefont {A.}~\bibnamefont {Butcher}}, \bibinfo {author} {\bibfnamefont {R.}~\bibnamefont {Kirk}}, \bibinfo {author} {\bibfnamefont {J.}~\bibnamefont {Monroe}}, \ and\ \bibinfo {author} {\bibfnamefont {S.~M.}\ \bibnamefont {West}},\ }\href {\doibase 10.1088/1475-7516/2017/10/035} {\bibfield  {journal} {\bibinfo  {journal} {JCAP}\ }\textbf {\bibinfo {volume} {10}},\ \bibinfo {pages} {035} (\bibinfo {year} {2017})},\ \Eprint {http://arxiv.org/abs/1610.01840} {arXiv:1610.01840 [hep-ph]} \BibitemShut {NoStop}%
\bibitem [{\citenamefont {Coskuner}\ \emph {et~al.}(2019)\citenamefont {Coskuner}, \citenamefont {Grabowska}, \citenamefont {Knapen},\ and\ \citenamefont {Zurek}}]{Models:Nuggets}%
  \BibitemOpen
  \bibfield  {author} {\bibinfo {author} {\bibfnamefont {A.}~\bibnamefont {Coskuner}}, \bibinfo {author} {\bibfnamefont {D.~M.}\ \bibnamefont {Grabowska}}, \bibinfo {author} {\bibfnamefont {S.}~\bibnamefont {Knapen}}, \ and\ \bibinfo {author} {\bibfnamefont {K.~M.}\ \bibnamefont {Zurek}},\ }\href {\doibase 10.1103/PhysRevD.100.035025} {\bibfield  {journal} {\bibinfo  {journal} {Phys. Rev.}\ }\textbf {\bibinfo {volume} {D100}},\ \bibinfo {pages} {035025} (\bibinfo {year} {2019})},\ \Eprint {http://arxiv.org/abs/1812.07573} {arXiv:1812.07573 [hep-ph]} \BibitemShut {NoStop}%
\bibitem [{\citenamefont {Bramante}\ \emph {et~al.}(2019{\natexlab{a}})\citenamefont {Bramante}, \citenamefont {Broerman}, \citenamefont {Kumar}, \citenamefont {Lang}, \citenamefont {Pospelov},\ and\ \citenamefont {Raj}}]{Bramante:2018tos}%
  \BibitemOpen
  \bibfield  {author} {\bibinfo {author} {\bibfnamefont {J.}~\bibnamefont {Bramante}}, \bibinfo {author} {\bibfnamefont {B.}~\bibnamefont {Broerman}}, \bibinfo {author} {\bibfnamefont {J.}~\bibnamefont {Kumar}}, \bibinfo {author} {\bibfnamefont {R.~F.}\ \bibnamefont {Lang}}, \bibinfo {author} {\bibfnamefont {M.}~\bibnamefont {Pospelov}}, \ and\ \bibinfo {author} {\bibfnamefont {N.}~\bibnamefont {Raj}},\ }\href {\doibase 10.1103/PhysRevD.99.083010} {\bibfield  {journal} {\bibinfo  {journal} {Phys. Rev. D}\ }\textbf {\bibinfo {volume} {99}},\ \bibinfo {pages} {083010} (\bibinfo {year} {2019}{\natexlab{a}})},\ \Eprint {http://arxiv.org/abs/1812.09325} {arXiv:1812.09325 [hep-ph]} \BibitemShut {NoStop}%
\bibitem [{\citenamefont {Bai}\ \emph {et~al.}(2020)\citenamefont {Bai}, \citenamefont {Korwar},\ and\ \citenamefont {Orlofsky}}]{Models:EWSymMonopoles:Bai:2020ttp}%
  \BibitemOpen
  \bibfield  {author} {\bibinfo {author} {\bibfnamefont {Y.}~\bibnamefont {Bai}}, \bibinfo {author} {\bibfnamefont {M.}~\bibnamefont {Korwar}}, \ and\ \bibinfo {author} {\bibfnamefont {N.}~\bibnamefont {Orlofsky}},\ }\href {\doibase 10.1007/JHEP07(2020)167} {\bibfield  {journal} {\bibinfo  {journal} {JHEP}\ }\textbf {\bibinfo {volume} {07}},\ \bibinfo {pages} {167} (\bibinfo {year} {2020})},\ \Eprint {http://arxiv.org/abs/2005.00503} {arXiv:2005.00503 [hep-ph]} \BibitemShut {NoStop}%
\bibitem [{\citenamefont {Ponton}\ \emph {et~al.}(2019)\citenamefont {Ponton}, \citenamefont {Bai},\ and\ \citenamefont {Jain}}]{Models:ElectroweakBalls}%
  \BibitemOpen
  \bibfield  {author} {\bibinfo {author} {\bibfnamefont {E.}~\bibnamefont {Ponton}}, \bibinfo {author} {\bibfnamefont {Y.}~\bibnamefont {Bai}}, \ and\ \bibinfo {author} {\bibfnamefont {B.}~\bibnamefont {Jain}},\ }\href@noop {} {\  (\bibinfo {year} {2019})},\ \Eprint {http://arxiv.org/abs/1906.10739} {arXiv:1906.10739 [hep-ph]} \BibitemShut {NoStop}%
\bibitem [{\citenamefont {Bai}\ and\ \citenamefont {Orlofsky}(2020)}]{Models:PlanckScaleBHRelicsBaiOrlofsky2019}%
  \BibitemOpen
  \bibfield  {author} {\bibinfo {author} {\bibfnamefont {Y.}~\bibnamefont {Bai}}\ and\ \bibinfo {author} {\bibfnamefont {N.}~\bibnamefont {Orlofsky}},\ }\href {\doibase 10.1103/PhysRevD.101.055006} {\bibfield  {journal} {\bibinfo  {journal} {Phys. Rev. D}\ }\textbf {\bibinfo {volume} {101}},\ \bibinfo {pages} {055006} (\bibinfo {year} {2020})},\ \Eprint {http://arxiv.org/abs/1906.04858} {arXiv:1906.04858 [hep-ph]} \BibitemShut {NoStop}%
\bibitem [{\citenamefont {Lehmann}\ \emph {et~al.}(2019)\citenamefont {Lehmann}, \citenamefont {Johnson}, \citenamefont {Profumo},\ and\ \citenamefont {Schwemberger}}]{Models:PlanckScaleBHRelicsSantaCruz}%
  \BibitemOpen
  \bibfield  {author} {\bibinfo {author} {\bibfnamefont {B.~V.}\ \bibnamefont {Lehmann}}, \bibinfo {author} {\bibfnamefont {C.}~\bibnamefont {Johnson}}, \bibinfo {author} {\bibfnamefont {S.}~\bibnamefont {Profumo}}, \ and\ \bibinfo {author} {\bibfnamefont {T.}~\bibnamefont {Schwemberger}},\ }\href@noop {} {\  (\bibinfo {year} {2019})},\ \Eprint {http://arxiv.org/abs/1906.06348} {arXiv:1906.06348 [hep-ph]} \BibitemShut {NoStop}%
\bibitem [{\citenamefont {Bernabei}\ \emph {et~al.}(1999)\citenamefont {Bernabei}, \citenamefont {Belli}, \citenamefont {Cerulli}, \citenamefont {Montecchia}, \citenamefont {Amato}, \citenamefont {Ignesti}, \citenamefont {Incicchitti}, \citenamefont {Prosperi}, \citenamefont {Dai}, \citenamefont {He}, \citenamefont {Kuang}, \citenamefont {Ma}, \citenamefont {Sun},\ and\ \citenamefont {Ye}}]{DAMA1999}%
  \BibitemOpen
  \bibfield  {author} {\bibinfo {author} {\bibfnamefont {R.}~\bibnamefont {Bernabei}}, \bibinfo {author} {\bibfnamefont {P.}~\bibnamefont {Belli}}, \bibinfo {author} {\bibfnamefont {R.}~\bibnamefont {Cerulli}}, \bibinfo {author} {\bibfnamefont {F.}~\bibnamefont {Montecchia}}, \bibinfo {author} {\bibfnamefont {M.}~\bibnamefont {Amato}}, \bibinfo {author} {\bibfnamefont {G.}~\bibnamefont {Ignesti}}, \bibinfo {author} {\bibfnamefont {A.}~\bibnamefont {Incicchitti}}, \bibinfo {author} {\bibfnamefont {D.}~\bibnamefont {Prosperi}}, \bibinfo {author} {\bibfnamefont {C.~J.}\ \bibnamefont {Dai}}, \bibinfo {author} {\bibfnamefont {H.~L.}\ \bibnamefont {He}}, \bibinfo {author} {\bibfnamefont {H.~H.}\ \bibnamefont {Kuang}}, \bibinfo {author} {\bibfnamefont {J.~M.}\ \bibnamefont {Ma}}, \bibinfo {author} {\bibfnamefont {G.~X.}\ \bibnamefont {Sun}}, \ and\ \bibinfo {author} {\bibfnamefont {Z.}~\bibnamefont {Ye}},\ }\href {\doibase 10.1103/PhysRevLett.83.4918} {\bibfield  {journal} {\bibinfo  {journal} {Phys. Rev. Lett.}\
  }\textbf {\bibinfo {volume} {83}},\ \bibinfo {pages} {4918} (\bibinfo {year} {1999})}\BibitemShut {NoStop}%
\bibitem [{\citenamefont {Albuquerque}\ and\ \citenamefont {Baudis}(2003)}]{EdelweissCDMSAlbuquerqueBaudis2003}%
  \BibitemOpen
  \bibfield  {author} {\bibinfo {author} {\bibfnamefont {I.~F.~M.}\ \bibnamefont {Albuquerque}}\ and\ \bibinfo {author} {\bibfnamefont {L.}~\bibnamefont {Baudis}},\ }\href {\doibase 10.1103/PhysRevLett.90.221301} {\bibfield  {journal} {\bibinfo  {journal} {Phys. Rev. Lett.}\ }\textbf {\bibinfo {volume} {90}},\ \bibinfo {pages} {221301} (\bibinfo {year} {2003})}\BibitemShut {NoStop}%
\bibitem [{\citenamefont {Kavanagh}(2018)}]{KavanaghSuperheavy:2017cru}%
  \BibitemOpen
  \bibfield  {author} {\bibinfo {author} {\bibfnamefont {B.~J.}\ \bibnamefont {Kavanagh}},\ }\href {\doibase 10.1103/PhysRevD.97.123013} {\bibfield  {journal} {\bibinfo  {journal} {Phys. Rev. D}\ }\textbf {\bibinfo {volume} {97}},\ \bibinfo {pages} {123013} (\bibinfo {year} {2018})},\ \Eprint {http://arxiv.org/abs/1712.04901} {arXiv:1712.04901 [hep-ph]} \BibitemShut {NoStop}%
\bibitem [{\citenamefont {Broerman}(2022)}]{PICO:MS:Broerman2022}%
  \BibitemOpen
  \bibfield  {author} {\bibinfo {author} {\bibfnamefont {B.}~\bibnamefont {Broerman}},\ }\emph {\bibinfo {title} {{New ideas for tonne-scale bubble chambers and a search for superheavy dark matter with PICO-60}}},\ \href@noop {} {Ph.D. thesis},\ \bibinfo  {school} {Queen's U., Kingston} (\bibinfo {year} {2022})\BibitemShut {NoStop}%
\bibitem [{\citenamefont {Clark}\ \emph {et~al.}(2020)\citenamefont {Clark}, \citenamefont {Depoian}, \citenamefont {Elshimy}, \citenamefont {Kopec}, \citenamefont {Lang}, \citenamefont {Li},\ and\ \citenamefont {Qin}}]{XENON1T:MSSSprojexn:Clark:2020mna}%
  \BibitemOpen
  \bibfield  {author} {\bibinfo {author} {\bibfnamefont {M.}~\bibnamefont {Clark}}, \bibinfo {author} {\bibfnamefont {A.}~\bibnamefont {Depoian}}, \bibinfo {author} {\bibfnamefont {B.}~\bibnamefont {Elshimy}}, \bibinfo {author} {\bibfnamefont {A.}~\bibnamefont {Kopec}}, \bibinfo {author} {\bibfnamefont {R.~F.}\ \bibnamefont {Lang}}, \bibinfo {author} {\bibfnamefont {S.}~\bibnamefont {Li}}, \ and\ \bibinfo {author} {\bibfnamefont {J.}~\bibnamefont {Qin}},\ }\href {\doibase 10.1103/PhysRevD.102.123026} {\bibfield  {journal} {\bibinfo  {journal} {Phys. Rev. D}\ }\textbf {\bibinfo {volume} {102}},\ \bibinfo {pages} {123026} (\bibinfo {year} {2020})},\ \Eprint {http://arxiv.org/abs/2009.07909} {arXiv:2009.07909 [hep-ph]} \BibitemShut {NoStop}%
\bibitem [{\citenamefont {Bhoonah}\ \emph {et~al.}(2021)\citenamefont {Bhoonah}, \citenamefont {Bramante}, \citenamefont {Courtman},\ and\ \citenamefont {Song}}]{PlasticEtch:Bhoonah2020fys}%
  \BibitemOpen
  \bibfield  {author} {\bibinfo {author} {\bibfnamefont {A.}~\bibnamefont {Bhoonah}}, \bibinfo {author} {\bibfnamefont {J.}~\bibnamefont {Bramante}}, \bibinfo {author} {\bibfnamefont {B.}~\bibnamefont {Courtman}}, \ and\ \bibinfo {author} {\bibfnamefont {N.}~\bibnamefont {Song}},\ }\href {\doibase 10.1103/PhysRevD.103.103001} {\bibfield  {journal} {\bibinfo  {journal} {Phys. Rev. D}\ }\textbf {\bibinfo {volume} {103}},\ \bibinfo {pages} {103001} (\bibinfo {year} {2021})},\ \Eprint {http://arxiv.org/abs/2012.13406} {arXiv:2012.13406 [hep-ph]} \BibitemShut {NoStop}%
\bibitem [{\citenamefont {Cappiello}\ \emph {et~al.}(2021)\citenamefont {Cappiello}, \citenamefont {Collar},\ and\ \citenamefont {Beacom}}]{CollarBeacomCappiello2021}%
  \BibitemOpen
  \bibfield  {author} {\bibinfo {author} {\bibfnamefont {C.~V.}\ \bibnamefont {Cappiello}}, \bibinfo {author} {\bibfnamefont {J.~I.}\ \bibnamefont {Collar}}, \ and\ \bibinfo {author} {\bibfnamefont {J.~F.}\ \bibnamefont {Beacom}},\ }\href {\doibase 10.1103/PhysRevD.103.023019} {\bibfield  {journal} {\bibinfo  {journal} {Phys. Rev. D}\ }\textbf {\bibinfo {volume} {103}},\ \bibinfo {pages} {023019} (\bibinfo {year} {2021})}\BibitemShut {NoStop}%
\bibitem [{\citenamefont {Adhikari}\ \emph {et~al.}(2022)\citenamefont {Adhikari} \emph {et~al.}}]{DEAP:MS:2021raj}%
  \BibitemOpen
  \bibfield  {author} {\bibinfo {author} {\bibfnamefont {P.}~\bibnamefont {Adhikari}} \emph {et~al.} (\bibinfo {collaboration} {(DEAP Collaboration)\textdaggerdbl{}, DEAP}),\ }\href {\doibase 10.1103/PhysRevLett.128.011801} {\bibfield  {journal} {\bibinfo  {journal} {Phys. Rev. Lett.}\ }\textbf {\bibinfo {volume} {128}},\ \bibinfo {pages} {011801} (\bibinfo {year} {2022})},\ \Eprint {http://arxiv.org/abs/2108.09405} {arXiv:2108.09405 [astro-ph.CO]} \BibitemShut {NoStop}%
\bibitem [{\citenamefont {Aprile}\ \emph {et~al.}(2023)\citenamefont {Aprile} \emph {et~al.}}]{XENON1T:MS:2023iku}%
  \BibitemOpen
  \bibfield  {author} {\bibinfo {author} {\bibfnamefont {E.}~\bibnamefont {Aprile}} \emph {et~al.} (\bibinfo {collaboration} {XENON}),\ }\href {\doibase 10.1103/PhysRevLett.130.261002} {\bibfield  {journal} {\bibinfo  {journal} {Phys. Rev. Lett.}\ }\textbf {\bibinfo {volume} {130}},\ \bibinfo {pages} {261002} (\bibinfo {year} {2023})},\ \Eprint {http://arxiv.org/abs/2304.10931} {arXiv:2304.10931 [hep-ex]} \BibitemShut {NoStop}%
\bibitem [{\citenamefont {Aalbers}\ \emph {et~al.}(2024{\natexlab{a}})\citenamefont {Aalbers} \emph {et~al.}}]{LZ:MS:2024psa}%
  \BibitemOpen
  \bibfield  {author} {\bibinfo {author} {\bibfnamefont {J.}~\bibnamefont {Aalbers}} \emph {et~al.} (\bibinfo {collaboration} {LZ}),\ }\href@noop {} {\  (\bibinfo {year} {2024}{\natexlab{a}})},\ \Eprint {http://arxiv.org/abs/2402.08865} {arXiv:2402.08865 [hep-ex]} \BibitemShut {NoStop}%
\bibitem [{\citenamefont {Raj}\ and\ \citenamefont {Mondal}(2024)}]{Raj:2024guv}%
  \BibitemOpen
  \bibfield  {author} {\bibinfo {author} {\bibfnamefont {N.}~\bibnamefont {Raj}}\ and\ \bibinfo {author} {\bibfnamefont {B.}~\bibnamefont {Mondal}},\ }\href@noop {} {\  (\bibinfo {year} {2024})},\ \Eprint {http://arxiv.org/abs/2406.17015} {arXiv:2406.17015 [hep-ph]} \BibitemShut {NoStop}%
\bibitem [{\citenamefont {Bramante}\ \emph {et~al.}(2018)\citenamefont {Bramante}, \citenamefont {Broerman}, \citenamefont {Lang},\ and\ \citenamefont {Raj}}]{Bramante:2018qbc}%
  \BibitemOpen
  \bibfield  {author} {\bibinfo {author} {\bibfnamefont {J.}~\bibnamefont {Bramante}}, \bibinfo {author} {\bibfnamefont {B.}~\bibnamefont {Broerman}}, \bibinfo {author} {\bibfnamefont {R.~F.}\ \bibnamefont {Lang}}, \ and\ \bibinfo {author} {\bibfnamefont {N.}~\bibnamefont {Raj}},\ }\href {\doibase 10.1103/PhysRevD.98.083516} {\bibfield  {journal} {\bibinfo  {journal} {Phys. Rev. D}\ }\textbf {\bibinfo {volume} {98}},\ \bibinfo {pages} {083516} (\bibinfo {year} {2018})},\ \Eprint {http://arxiv.org/abs/1803.08044} {arXiv:1803.08044 [hep-ph]} \BibitemShut {NoStop}%
\bibitem [{\citenamefont {Kavanagh}\ \emph {et~al.}(2021)\citenamefont {Kavanagh}, \citenamefont {Emken},\ and\ \citenamefont {Catena}}]{Kavanagh:2020cvn}%
  \BibitemOpen
  \bibfield  {author} {\bibinfo {author} {\bibfnamefont {B.~J.}\ \bibnamefont {Kavanagh}}, \bibinfo {author} {\bibfnamefont {T.}~\bibnamefont {Emken}}, \ and\ \bibinfo {author} {\bibfnamefont {R.}~\bibnamefont {Catena}},\ }\href {\doibase 10.1103/PhysRevD.104.083023} {\bibfield  {journal} {\bibinfo  {journal} {Phys. Rev. D}\ }\textbf {\bibinfo {volume} {104}},\ \bibinfo {pages} {083023} (\bibinfo {year} {2021})},\ \Eprint {http://arxiv.org/abs/2004.01621} {arXiv:2004.01621 [hep-ph]} \BibitemShut {NoStop}%
\bibitem [{\citenamefont {Bramante}\ \emph {et~al.}(2019{\natexlab{b}})\citenamefont {Bramante}, \citenamefont {Kumar},\ and\ \citenamefont {Raj}}]{Bramante:2019yss}%
  \BibitemOpen
  \bibfield  {author} {\bibinfo {author} {\bibfnamefont {J.}~\bibnamefont {Bramante}}, \bibinfo {author} {\bibfnamefont {J.}~\bibnamefont {Kumar}}, \ and\ \bibinfo {author} {\bibfnamefont {N.}~\bibnamefont {Raj}},\ }\href {\doibase 10.1103/PhysRevD.100.123016} {\bibfield  {journal} {\bibinfo  {journal} {Phys. Rev. D}\ }\textbf {\bibinfo {volume} {100}},\ \bibinfo {pages} {123016} (\bibinfo {year} {2019}{\natexlab{b}})},\ \Eprint {http://arxiv.org/abs/1910.05380} {arXiv:1910.05380 [hep-ph]} \BibitemShut {NoStop}%
\bibitem [{\citenamefont {Vahsen}\ \emph {et~al.}(2021)\citenamefont {Vahsen}, \citenamefont {O'Hare},\ and\ \citenamefont {Loomba}}]{reviewDirectional:Vahsen:2021gnb}%
  \BibitemOpen
  \bibfield  {author} {\bibinfo {author} {\bibfnamefont {S.~E.}\ \bibnamefont {Vahsen}}, \bibinfo {author} {\bibfnamefont {C.~A.~J.}\ \bibnamefont {O'Hare}}, \ and\ \bibinfo {author} {\bibfnamefont {D.}~\bibnamefont {Loomba}},\ }\href {\doibase 10.1146/annurev-nucl-020821-035016} {\bibfield  {journal} {\bibinfo  {journal} {Ann. Rev. Nucl. Part. Sci.}\ }\textbf {\bibinfo {volume} {71}},\ \bibinfo {pages} {189} (\bibinfo {year} {2021})},\ \Eprint {http://arxiv.org/abs/2102.04596} {arXiv:2102.04596 [physics.ins-det]} \BibitemShut {NoStop}%
\bibitem [{\citenamefont {Church}\ \emph {et~al.}(2020)\citenamefont {Church}, \citenamefont {Jackson},\ and\ \citenamefont {Saldanha}}]{DUNEModuleDM:PNL2020}%
  \BibitemOpen
  \bibfield  {author} {\bibinfo {author} {\bibfnamefont {E.}~\bibnamefont {Church}}, \bibinfo {author} {\bibfnamefont {C.~M.}\ \bibnamefont {Jackson}}, \ and\ \bibinfo {author} {\bibfnamefont {R.}~\bibnamefont {Saldanha}},\ }\href {\doibase 10.1088/1748-0221/15/09/P09026} {\bibfield  {journal} {\bibinfo  {journal} {JINST}\ }\textbf {\bibinfo {volume} {15}},\ \bibinfo {pages} {P09026} (\bibinfo {year} {2020})},\ \Eprint {http://arxiv.org/abs/2005.04824} {arXiv:2005.04824 [physics.ins-det]} \BibitemShut {NoStop}%
\bibitem [{\citenamefont {Avasthi}\ \emph {et~al.}(2022)\citenamefont {Avasthi} \emph {et~al.}}]{DUNEModuleDM:snowmass:Avasthi2022}%
  \BibitemOpen
  \bibfield  {author} {\bibinfo {author} {\bibfnamefont {A.}~\bibnamefont {Avasthi}} \emph {et~al.},\ }in\ \href@noop {} {\emph {\bibinfo {booktitle} {{Snowmass 2021}}}}\ (\bibinfo {year} {2022})\ \Eprint {http://arxiv.org/abs/2203.08821} {arXiv:2203.08821 [physics.ins-det]} \BibitemShut {NoStop}%
\bibitem [{\citenamefont {Bezerra}\ \emph {et~al.}(2023)\citenamefont {Bezerra} \emph {et~al.}}]{DUNEModuleDM:Bezerra2023}%
  \BibitemOpen
  \bibfield  {author} {\bibinfo {author} {\bibfnamefont {T.}~\bibnamefont {Bezerra}} \emph {et~al.},\ }\href {\doibase 10.1088/1361-6471/acc394} {\bibfield  {journal} {\bibinfo  {journal} {J. Phys. G}\ }\textbf {\bibinfo {volume} {50}},\ \bibinfo {pages} {060502} (\bibinfo {year} {2023})},\ \Eprint {http://arxiv.org/abs/2301.11878} {arXiv:2301.11878 [hep-ex]} \BibitemShut {NoStop}%
\bibitem [{\citenamefont {Roberts}\ \emph {et~al.}(2019)\citenamefont {Roberts} \emph {et~al.}}]{ARIADNE:Roberts:2018sww}%
  \BibitemOpen
  \bibfield  {author} {\bibinfo {author} {\bibfnamefont {A.}~\bibnamefont {Roberts}} \emph {et~al.},\ }\href {\doibase 10.1088/1748-0221/14/06/P06001} {\bibfield  {journal} {\bibinfo  {journal} {JINST}\ }\textbf {\bibinfo {volume} {14}},\ \bibinfo {pages} {P06001} (\bibinfo {year} {2019})},\ \Eprint {http://arxiv.org/abs/1810.09955} {arXiv:1810.09955 [physics.ins-det]} \BibitemShut {NoStop}%
\bibitem [{\citenamefont {Avasthi}\ \emph {et~al.}(2021)\citenamefont {Avasthi} \emph {et~al.}}]{Xekton:Avasthi:2021lgy}%
  \BibitemOpen
  \bibfield  {author} {\bibinfo {author} {\bibfnamefont {A.}~\bibnamefont {Avasthi}} \emph {et~al.},\ }\href {\doibase 10.1103/PhysRevD.104.112007} {\bibfield  {journal} {\bibinfo  {journal} {Phys. Rev. D}\ }\textbf {\bibinfo {volume} {104}},\ \bibinfo {pages} {112007} (\bibinfo {year} {2021})},\ \Eprint {http://arxiv.org/abs/2110.01537} {arXiv:2110.01537 [physics.ins-det]} \BibitemShut {NoStop}%
\bibitem [{\citenamefont {Anker}\ \emph {et~al.}(2024)\citenamefont {Anker} \emph {et~al.}}]{Xekton:Anker:2024xfz}%
  \BibitemOpen
  \bibfield  {author} {\bibinfo {author} {\bibfnamefont {A.}~\bibnamefont {Anker}} \emph {et~al.},\ }in\ \href@noop {} {\emph {\bibinfo {booktitle} {{Workshop on Xenon Detector 0\ensuremath{\nu}\ensuremath{\beta}\ensuremath{\beta} Searches: Steps Towards the Kilotonne Scale}}}}\ (\bibinfo {year} {2024})\ \Eprint {http://arxiv.org/abs/2404.19050} {arXiv:2404.19050 [nucl-ex]} \BibitemShut {NoStop}%
\bibitem [{\citenamefont {Vahsen}\ \emph {et~al.}(2020)\citenamefont {Vahsen} \emph {et~al.}}]{CYGNUS:2020pzb}%
  \BibitemOpen
  \bibfield  {author} {\bibinfo {author} {\bibfnamefont {S.~E.}\ \bibnamefont {Vahsen}} \emph {et~al.},\ }\href@noop {} {\  (\bibinfo {year} {2020})},\ \Eprint {http://arxiv.org/abs/2008.12587} {arXiv:2008.12587 [physics.ins-det]} \BibitemShut {NoStop}%
\bibitem [{\citenamefont {Macolino}(2020)}]{DARWIN:Macolino:2020uqq}%
  \BibitemOpen
  \bibfield  {author} {\bibinfo {author} {\bibfnamefont {C.}~\bibnamefont {Macolino}} (\bibinfo {collaboration} {DARWIN}),\ }\href {\doibase 10.1088/1742-6596/1468/1/012068} {\bibfield  {journal} {\bibinfo  {journal} {J. Phys. Conf. Ser.}\ }\textbf {\bibinfo {volume} {1468}},\ \bibinfo {pages} {012068} (\bibinfo {year} {2020})}\BibitemShut {NoStop}%
\bibitem [{\citenamefont {Aalbers}\ \emph {et~al.}(2024{\natexlab{b}})\citenamefont {Aalbers} \emph {et~al.}}]{XLZD:2024gxx}%
  \BibitemOpen
  \bibfield  {author} {\bibinfo {author} {\bibfnamefont {J.}~\bibnamefont {Aalbers}} \emph {et~al.} (\bibinfo {collaboration} {XLZD}),\ }\href@noop {} {\  (\bibinfo {year} {2024}{\natexlab{b}})},\ \Eprint {http://arxiv.org/abs/2410.17137} {arXiv:2410.17137 [hep-ex]} \BibitemShut {NoStop}%
\bibitem [{\citenamefont {Galbiati}\ \emph {et~al.}(2018)\citenamefont {Galbiati} \emph {et~al.}}]{ARGO:2018}%
  \BibitemOpen
  \bibfield  {author} {\bibinfo {author} {\bibfnamefont {C.}~\bibnamefont {Galbiati}} \emph {et~al.} (\bibinfo {collaboration} {GADMC}),\ }\href {https://indico.cern.ch/event/765096/contributions/3295671/attachments/1785196/2906164/DarkSide-Argo_ESPP_Dec_17_2017.pdf} {\bibfield  {journal} {\bibinfo  {journal} {Input to the European Particle Physics Strategy Update 2018-2020}\ } (\bibinfo {year} {2018})}\BibitemShut {NoStop}%
\bibitem [{ARG()}]{ARGOSnowmassLOI}%
  \BibitemOpen
  \href@noop {} {}\bibinfo {howpublished} {\url{https://www.snowmass21.org/docs/files/summaries/CF/SNOWMASS21-CF1_CF0_Giovanetti-172.pdf}}\BibitemShut {NoStop}%
\bibitem [{\citenamefont {Myeong}\ \emph {et~al.}(2018{\natexlab{a}})\citenamefont {Myeong}, \citenamefont {Evans}, \citenamefont {Belokurov}, \citenamefont {Amorisco},\ and\ \citenamefont {Koposov}}]{streamSDSSGaia2017}%
  \BibitemOpen
  \bibfield  {author} {\bibinfo {author} {\bibfnamefont {G.~C.}\ \bibnamefont {Myeong}}, \bibinfo {author} {\bibfnamefont {N.~W.}\ \bibnamefont {Evans}}, \bibinfo {author} {\bibfnamefont {V.}~\bibnamefont {Belokurov}}, \bibinfo {author} {\bibfnamefont {N.~C.}\ \bibnamefont {Amorisco}}, \ and\ \bibinfo {author} {\bibfnamefont {S.}~\bibnamefont {Koposov}},\ }\href {\doibase 10.1093/mnras/stx3262} {\bibfield  {journal} {\bibinfo  {journal} {Mon. Not. Roy. Astron. Soc.}\ }\textbf {\bibinfo {volume} {475}},\ \bibinfo {pages} {1537} (\bibinfo {year} {2018}{\natexlab{a}})},\ \Eprint {http://arxiv.org/abs/1712.04071} {arXiv:1712.04071 [astro-ph.GA]} \BibitemShut {NoStop}%
\bibitem [{\citenamefont {{Belokurov}}\ \emph {et~al.}(2018)\citenamefont {{Belokurov}}, \citenamefont {{Erkal}}, \citenamefont {{Evans}}, \citenamefont {{Koposov}},\ and\ \citenamefont {{Deason}}}]{streamGaias2018}%
  \BibitemOpen
  \bibfield  {author} {\bibinfo {author} {\bibfnamefont {V.}~\bibnamefont {{Belokurov}}}, \bibinfo {author} {\bibfnamefont {D.}~\bibnamefont {{Erkal}}}, \bibinfo {author} {\bibfnamefont {N.~W.}\ \bibnamefont {{Evans}}}, \bibinfo {author} {\bibfnamefont {S.~E.}\ \bibnamefont {{Koposov}}}, \ and\ \bibinfo {author} {\bibfnamefont {A.~J.}\ \bibnamefont {{Deason}}},\ }\href {\doibase 10.1093/mnras/sty982} {\bibfield  {journal} {\bibinfo  {journal} {MNRAS}\ }\textbf {\bibinfo {volume} {478}},\ \bibinfo {pages} {611} (\bibinfo {year} {2018})},\ \Eprint {http://arxiv.org/abs/1802.03414} {arXiv:1802.03414 [astro-ph.GA]} \BibitemShut {NoStop}%
\bibitem [{\citenamefont {{Naidu}}\ \emph {et~al.}(2020)\citenamefont {{Naidu}}, \citenamefont {{Conroy}}, \citenamefont {{Bonaca}}, \citenamefont {{Johnson}}, \citenamefont {{Ting}}, \citenamefont {{Caldwell}}, \citenamefont {{Zaritsky}},\ and\ \citenamefont {{Cargile}}}]{streamGaia2020}%
  \BibitemOpen
  \bibfield  {author} {\bibinfo {author} {\bibfnamefont {R.~P.}\ \bibnamefont {{Naidu}}}, \bibinfo {author} {\bibfnamefont {C.}~\bibnamefont {{Conroy}}}, \bibinfo {author} {\bibfnamefont {A.}~\bibnamefont {{Bonaca}}}, \bibinfo {author} {\bibfnamefont {B.~D.}\ \bibnamefont {{Johnson}}}, \bibinfo {author} {\bibfnamefont {Y.-S.}\ \bibnamefont {{Ting}}}, \bibinfo {author} {\bibfnamefont {N.}~\bibnamefont {{Caldwell}}}, \bibinfo {author} {\bibfnamefont {D.}~\bibnamefont {{Zaritsky}}}, \ and\ \bibinfo {author} {\bibfnamefont {P.~A.}\ \bibnamefont {{Cargile}}},\ }\href {\doibase 10.3847/1538-4357/abaef4} {\bibfield  {journal} {\bibinfo  {journal} {\apj}\ }\textbf {\bibinfo {volume} {901}},\ \bibinfo {eid} {48} (\bibinfo {year} {2020})},\ \Eprint {http://arxiv.org/abs/2006.08625} {arXiv:2006.08625 [astro-ph.GA]} \BibitemShut {NoStop}%
\bibitem [{\citenamefont {{Dodd}}\ \emph {et~al.}(2023)\citenamefont {{Dodd}}, \citenamefont {{Callingham}}, \citenamefont {{Helmi}}, \citenamefont {{Matsuno}}, \citenamefont {{Ruiz-Lara}}, \citenamefont {{Balbinot}},\ and\ \citenamefont {{L{\"o}vdal}}}]{streamGaiaDR32022}%
  \BibitemOpen
  \bibfield  {author} {\bibinfo {author} {\bibfnamefont {E.}~\bibnamefont {{Dodd}}}, \bibinfo {author} {\bibfnamefont {T.~M.}\ \bibnamefont {{Callingham}}}, \bibinfo {author} {\bibfnamefont {A.}~\bibnamefont {{Helmi}}}, \bibinfo {author} {\bibfnamefont {T.}~\bibnamefont {{Matsuno}}}, \bibinfo {author} {\bibfnamefont {T.}~\bibnamefont {{Ruiz-Lara}}}, \bibinfo {author} {\bibfnamefont {E.}~\bibnamefont {{Balbinot}}}, \ and\ \bibinfo {author} {\bibfnamefont {S.}~\bibnamefont {{L{\"o}vdal}}},\ }\href {\doibase 10.1051/0004-6361/202244546} {\bibfield  {journal} {\bibinfo  {journal} {AAP}\ }\textbf {\bibinfo {volume} {670}},\ \bibinfo {eid} {L2} (\bibinfo {year} {2023})},\ \Eprint {http://arxiv.org/abs/2206.11248} {arXiv:2206.11248 [astro-ph.GA]} \BibitemShut {NoStop}%
\bibitem [{\citenamefont {{Yuan}}\ \emph {et~al.}(2020)\citenamefont {{Yuan}}, \citenamefont {{Myeong}}, \citenamefont {{Beers}}, \citenamefont {{Evans}}, \citenamefont {{Lee}}, \citenamefont {{Banerjee}}, \citenamefont {{Gudin}}, \citenamefont {{Hattori}}, \citenamefont {{Li}}, \citenamefont {{Matsuno}}, \citenamefont {{Placco}}, \citenamefont {{Smith}}, \citenamefont {{Whitten}},\ and\ \citenamefont {{Zhao}}}]{streamLAMOST2023}%
  \BibitemOpen
  \bibfield  {author} {\bibinfo {author} {\bibfnamefont {Z.}~\bibnamefont {{Yuan}}}, \bibinfo {author} {\bibfnamefont {G.~C.}\ \bibnamefont {{Myeong}}}, \bibinfo {author} {\bibfnamefont {T.~C.}\ \bibnamefont {{Beers}}}, \bibinfo {author} {\bibfnamefont {N.~W.}\ \bibnamefont {{Evans}}}, \bibinfo {author} {\bibfnamefont {Y.~S.}\ \bibnamefont {{Lee}}}, \bibinfo {author} {\bibfnamefont {P.}~\bibnamefont {{Banerjee}}}, \bibinfo {author} {\bibfnamefont {D.}~\bibnamefont {{Gudin}}}, \bibinfo {author} {\bibfnamefont {K.}~\bibnamefont {{Hattori}}}, \bibinfo {author} {\bibfnamefont {H.}~\bibnamefont {{Li}}}, \bibinfo {author} {\bibfnamefont {T.}~\bibnamefont {{Matsuno}}}, \bibinfo {author} {\bibfnamefont {V.~M.}\ \bibnamefont {{Placco}}}, \bibinfo {author} {\bibfnamefont {M.~C.}\ \bibnamefont {{Smith}}}, \bibinfo {author} {\bibfnamefont {D.~D.}\ \bibnamefont {{Whitten}}}, \ and\ \bibinfo {author} {\bibfnamefont {G.}~\bibnamefont {{Zhao}}},\ }\href {\doibase 10.3847/1538-4357/ab6ef7} {\bibfield  {journal} {\bibinfo
  {journal} {\apj}\ }\textbf {\bibinfo {volume} {891}},\ \bibinfo {eid} {39} (\bibinfo {year} {2020})},\ \Eprint {http://arxiv.org/abs/1910.07538} {arXiv:1910.07538 [astro-ph.GA]} \BibitemShut {NoStop}%
\bibitem [{\citenamefont {Lewin}\ and\ \citenamefont {Smith}(1996)}]{Lewin:1995rx}%
  \BibitemOpen
  \bibfield  {author} {\bibinfo {author} {\bibfnamefont {J.~D.}\ \bibnamefont {Lewin}}\ and\ \bibinfo {author} {\bibfnamefont {P.~F.}\ \bibnamefont {Smith}},\ }\href {\doibase 10.1016/S0927-6505(96)00047-3} {\bibfield  {journal} {\bibinfo  {journal} {Astropart. Phys.}\ }\textbf {\bibinfo {volume} {6}},\ \bibinfo {pages} {87} (\bibinfo {year} {1996})}\BibitemShut {NoStop}%
\bibitem [{\citenamefont {Gaspert}\ \emph {et~al.}(2022)\citenamefont {Gaspert}, \citenamefont {Giampa},\ and\ \citenamefont {Morrissey}}]{nufloor:GaspertGiampaMorrissey:2021gyj}%
  \BibitemOpen
  \bibfield  {author} {\bibinfo {author} {\bibfnamefont {A.}~\bibnamefont {Gaspert}}, \bibinfo {author} {\bibfnamefont {P.}~\bibnamefont {Giampa}}, \ and\ \bibinfo {author} {\bibfnamefont {D.~E.}\ \bibnamefont {Morrissey}},\ }\href {\doibase 10.1103/PhysRevD.105.035020} {\bibfield  {journal} {\bibinfo  {journal} {Phys. Rev. D}\ }\textbf {\bibinfo {volume} {105}},\ \bibinfo {pages} {035020} (\bibinfo {year} {2022})},\ \Eprint {http://arxiv.org/abs/2108.03248} {arXiv:2108.03248 [hep-ph]} \BibitemShut {NoStop}%
\bibitem [{\citenamefont {Baxter}\ \emph {et~al.}(2021)\citenamefont {Baxter} \emph {et~al.}}]{Baxter:2021pqo}%
  \BibitemOpen
  \bibfield  {author} {\bibinfo {author} {\bibfnamefont {D.}~\bibnamefont {Baxter}} \emph {et~al.},\ }\href {\doibase 10.1140/epjc/s10052-021-09655-y} {\bibfield  {journal} {\bibinfo  {journal} {Eur. Phys. J. C}\ }\textbf {\bibinfo {volume} {81}},\ \bibinfo {pages} {907} (\bibinfo {year} {2021})},\ \Eprint {http://arxiv.org/abs/2105.00599} {arXiv:2105.00599 [hep-ex]} \BibitemShut {NoStop}%
\bibitem [{\citenamefont {Smith}\ and\ \citenamefont {Lewin}(1990)}]{SMITHLEWIN1990}%
  \BibitemOpen
  \bibfield  {author} {\bibinfo {author} {\bibfnamefont {P.}~\bibnamefont {Smith}}\ and\ \bibinfo {author} {\bibfnamefont {J.}~\bibnamefont {Lewin}},\ }\href {\doibase https://doi.org/10.1016/0370-1573(90)90081-C} {\bibfield  {journal} {\bibinfo  {journal} {Physics Reports}\ }\textbf {\bibinfo {volume} {187}},\ \bibinfo {pages} {203} (\bibinfo {year} {1990})}\BibitemShut {NoStop}%
\bibitem [{\citenamefont {Bozorgnia}\ \emph {et~al.}(2011)\citenamefont {Bozorgnia}, \citenamefont {Gelmini},\ and\ \citenamefont {Gondolo}}]{PhysRevD.84.023516}%
  \BibitemOpen
  \bibfield  {author} {\bibinfo {author} {\bibfnamefont {N.}~\bibnamefont {Bozorgnia}}, \bibinfo {author} {\bibfnamefont {G.~B.}\ \bibnamefont {Gelmini}}, \ and\ \bibinfo {author} {\bibfnamefont {P.}~\bibnamefont {Gondolo}},\ }\href {\doibase 10.1103/PhysRevD.84.023516} {\bibfield  {journal} {\bibinfo  {journal} {Phys. Rev. D}\ }\textbf {\bibinfo {volume} {84}},\ \bibinfo {pages} {023516} (\bibinfo {year} {2011})}\BibitemShut {NoStop}%
\bibitem [{\citenamefont {Mayet}\ \emph {et~al.}(2016)\citenamefont {Mayet}, \citenamefont {Green}, \citenamefont {Battat}, \citenamefont {Billard}, \citenamefont {Bozorgnia}, \citenamefont {Gelmini}, \citenamefont {Gondolo}, \citenamefont {Kavanagh}, \citenamefont {Lee}, \citenamefont {Loomba}, \citenamefont {Monroe}, \citenamefont {Morgan}, \citenamefont {O’Hare}, \citenamefont {Peter}, \citenamefont {Phan},\ and\ \citenamefont {Vahsen}}]{Mayet_2016}%
  \BibitemOpen
  \bibfield  {author} {\bibinfo {author} {\bibfnamefont {F.}~\bibnamefont {Mayet}}, \bibinfo {author} {\bibfnamefont {A.}~\bibnamefont {Green}}, \bibinfo {author} {\bibfnamefont {J.}~\bibnamefont {Battat}}, \bibinfo {author} {\bibfnamefont {J.}~\bibnamefont {Billard}}, \bibinfo {author} {\bibfnamefont {N.}~\bibnamefont {Bozorgnia}}, \bibinfo {author} {\bibfnamefont {G.}~\bibnamefont {Gelmini}}, \bibinfo {author} {\bibfnamefont {P.}~\bibnamefont {Gondolo}}, \bibinfo {author} {\bibfnamefont {B.}~\bibnamefont {Kavanagh}}, \bibinfo {author} {\bibfnamefont {S.}~\bibnamefont {Lee}}, \bibinfo {author} {\bibfnamefont {D.}~\bibnamefont {Loomba}}, \bibinfo {author} {\bibfnamefont {J.}~\bibnamefont {Monroe}}, \bibinfo {author} {\bibfnamefont {B.}~\bibnamefont {Morgan}}, \bibinfo {author} {\bibfnamefont {C.}~\bibnamefont {O’Hare}}, \bibinfo {author} {\bibfnamefont {A.}~\bibnamefont {Peter}}, \bibinfo {author} {\bibfnamefont {N.}~\bibnamefont {Phan}}, \ and\ \bibinfo {author} {\bibfnamefont {S.}~\bibnamefont {Vahsen}},\
  }\href {\doibase 10.1016/j.physrep.2016.02.007} {\bibfield  {journal} {\bibinfo  {journal} {Physics Reports}\ }\textbf {\bibinfo {volume} {627}},\ \bibinfo {pages} {1–49} (\bibinfo {year} {2016})}\BibitemShut {NoStop}%
\bibitem [{\citenamefont {Helm}(1956)}]{Helm:1956zz}%
  \BibitemOpen
  \bibfield  {author} {\bibinfo {author} {\bibfnamefont {R.~H.}\ \bibnamefont {Helm}},\ }\href {\doibase 10.1103/PhysRev.104.1466} {\bibfield  {journal} {\bibinfo  {journal} {Phys. Rev.}\ }\textbf {\bibinfo {volume} {104}},\ \bibinfo {pages} {1466} (\bibinfo {year} {1956})}\BibitemShut {NoStop}%
\bibitem [{\citenamefont {Digman}\ \emph {et~al.}(2019)\citenamefont {Digman}, \citenamefont {Cappiello}, \citenamefont {Beacom}, \citenamefont {Hirata},\ and\ \citenamefont {Peter}}]{Digman:2019wdm}%
  \BibitemOpen
  \bibfield  {author} {\bibinfo {author} {\bibfnamefont {M.~C.}\ \bibnamefont {Digman}}, \bibinfo {author} {\bibfnamefont {C.~V.}\ \bibnamefont {Cappiello}}, \bibinfo {author} {\bibfnamefont {J.~F.}\ \bibnamefont {Beacom}}, \bibinfo {author} {\bibfnamefont {C.~M.}\ \bibnamefont {Hirata}}, \ and\ \bibinfo {author} {\bibfnamefont {A.~H.~G.}\ \bibnamefont {Peter}},\ }\href {\doibase 10.1103/PhysRevD.100.063013} {\bibfield  {journal} {\bibinfo  {journal} {Phys. Rev. D}\ }\textbf {\bibinfo {volume} {100}},\ \bibinfo {pages} {063013} (\bibinfo {year} {2019})},\ \bibinfo {note} {[Erratum: Phys.Rev.D 106, 089902 (2022)]},\ \Eprint {http://arxiv.org/abs/1907.10618} {arXiv:1907.10618 [hep-ph]} \BibitemShut {NoStop}%
\bibitem [{\citenamefont {Feldman}\ and\ \citenamefont {Cousins}(1998)}]{Feldman:1997qc}%
  \BibitemOpen
  \bibfield  {author} {\bibinfo {author} {\bibfnamefont {G.~J.}\ \bibnamefont {Feldman}}\ and\ \bibinfo {author} {\bibfnamefont {R.~D.}\ \bibnamefont {Cousins}},\ }\href {\doibase 10.1103/PhysRevD.57.3873} {\bibfield  {journal} {\bibinfo  {journal} {Phys. Rev. D}\ }\textbf {\bibinfo {volume} {57}},\ \bibinfo {pages} {3873} (\bibinfo {year} {1998})},\ \Eprint {http://arxiv.org/abs/physics/9711021} {arXiv:physics/9711021} \BibitemShut {NoStop}%
\bibitem [{\citenamefont {Lewis}\ \emph {et~al.}(2015)\citenamefont {Lewis}, \citenamefont {Vahsen}, \citenamefont {Seong}, \citenamefont {Hedges}, \citenamefont {Jaegle},\ and\ \citenamefont {Thorpe}}]{Lewis_2015}%
  \BibitemOpen
  \bibfield  {author} {\bibinfo {author} {\bibfnamefont {P.}~\bibnamefont {Lewis}}, \bibinfo {author} {\bibfnamefont {S.}~\bibnamefont {Vahsen}}, \bibinfo {author} {\bibfnamefont {I.}~\bibnamefont {Seong}}, \bibinfo {author} {\bibfnamefont {M.}~\bibnamefont {Hedges}}, \bibinfo {author} {\bibfnamefont {I.}~\bibnamefont {Jaegle}}, \ and\ \bibinfo {author} {\bibfnamefont {T.}~\bibnamefont {Thorpe}},\ }\href {\doibase 10.1016/j.nima.2015.03.024} {\bibfield  {journal} {\bibinfo  {journal} {Nuclear Instruments and Methods in Physics Research Section A: Accelerators, Spectrometers, Detectors and Associated Equipment}\ }\textbf {\bibinfo {volume} {789}},\ \bibinfo {pages} {81–85} (\bibinfo {year} {2015})}\BibitemShut {NoStop}%
\bibitem [{\citenamefont {Bhoonah}\ \emph {et~al.}(2019)\citenamefont {Bhoonah}, \citenamefont {Bramante}, \citenamefont {Elahi},\ and\ \citenamefont {Schon}}]{gascloudcompendium:Bhoonah:2018gjb}%
  \BibitemOpen
  \bibfield  {author} {\bibinfo {author} {\bibfnamefont {A.}~\bibnamefont {Bhoonah}}, \bibinfo {author} {\bibfnamefont {J.}~\bibnamefont {Bramante}}, \bibinfo {author} {\bibfnamefont {F.}~\bibnamefont {Elahi}}, \ and\ \bibinfo {author} {\bibfnamefont {S.}~\bibnamefont {Schon}},\ }\href {\doibase 10.1103/PhysRevD.100.023001} {\bibfield  {journal} {\bibinfo  {journal} {Phys. Rev. D}\ }\textbf {\bibinfo {volume} {100}},\ \bibinfo {pages} {023001} (\bibinfo {year} {2019})},\ \Eprint {http://arxiv.org/abs/1812.10919} {arXiv:1812.10919 [hep-ph]} \BibitemShut {NoStop}%
\bibitem [{\citenamefont {Price}\ and\ \citenamefont {Salamon}(1986)}]{mica:pricesalamon1986}%
  \BibitemOpen
  \bibfield  {author} {\bibinfo {author} {\bibfnamefont {P.~B.}\ \bibnamefont {Price}}\ and\ \bibinfo {author} {\bibfnamefont {M.~H.}\ \bibnamefont {Salamon}},\ }\href {\doibase 10.1103/PhysRevLett.56.1226} {\bibfield  {journal} {\bibinfo  {journal} {Phys. Rev. Lett.}\ }\textbf {\bibinfo {volume} {56}},\ \bibinfo {pages} {1226} (\bibinfo {year} {1986})}\BibitemShut {NoStop}%
\bibitem [{\citenamefont {Segreto}(2021)}]{PhysRevD.103.043001}%
  \BibitemOpen
  \bibfield  {author} {\bibinfo {author} {\bibfnamefont {E.}~\bibnamefont {Segreto}},\ }\href {\doibase 10.1103/PhysRevD.103.043001} {\bibfield  {journal} {\bibinfo  {journal} {Phys. Rev. D}\ }\textbf {\bibinfo {volume} {103}},\ \bibinfo {pages} {043001} (\bibinfo {year} {2021})}\BibitemShut {NoStop}%
\bibitem [{LBi()}]{LBignell}%
  \BibitemOpen
  \href@noop {} {}\bibinfo {howpublished} {L. Bignell, private correspondence.}\BibitemShut {Stop}%
\bibitem [{\citenamefont {Helmi}\ and\ \citenamefont {White}(1999)}]{tidalstreamHelmi:1999ks}%
  \BibitemOpen
  \bibfield  {author} {\bibinfo {author} {\bibfnamefont {A.}~\bibnamefont {Helmi}}\ and\ \bibinfo {author} {\bibfnamefont {S.~D.~M.}\ \bibnamefont {White}},\ }\href {\doibase 10.1046/j.1365-8711.1999.02616.x} {\bibfield  {journal} {\bibinfo  {journal} {Mon. Not. Roy. Astron. Soc.}\ }\textbf {\bibinfo {volume} {307}},\ \bibinfo {pages} {495} (\bibinfo {year} {1999})},\ \Eprint {http://arxiv.org/abs/astro-ph/9901102} {arXiv:astro-ph/9901102} \BibitemShut {NoStop}%
\bibitem [{\citenamefont {Freese}\ \emph {et~al.}(2005)\citenamefont {Freese}, \citenamefont {Gondolo},\ and\ \citenamefont {Newberg}}]{tidalstreamFreese:2003tt}%
  \BibitemOpen
  \bibfield  {author} {\bibinfo {author} {\bibfnamefont {K.}~\bibnamefont {Freese}}, \bibinfo {author} {\bibfnamefont {P.}~\bibnamefont {Gondolo}}, \ and\ \bibinfo {author} {\bibfnamefont {H.~J.}\ \bibnamefont {Newberg}},\ }\href {\doibase 10.1103/PhysRevD.71.043516} {\bibfield  {journal} {\bibinfo  {journal} {Phys. Rev. D}\ }\textbf {\bibinfo {volume} {71}},\ \bibinfo {pages} {043516} (\bibinfo {year} {2005})},\ \Eprint {http://arxiv.org/abs/astro-ph/0309279} {arXiv:astro-ph/0309279} \BibitemShut {NoStop}%
\bibitem [{\citenamefont {Read}\ \emph {et~al.}(2009)\citenamefont {Read}, \citenamefont {Mayer}, \citenamefont {Brooks}, \citenamefont {Governato},\ and\ \citenamefont {Lake}}]{darkdiskRead2009}%
  \BibitemOpen
  \bibfield  {author} {\bibinfo {author} {\bibfnamefont {J.~I.}\ \bibnamefont {Read}}, \bibinfo {author} {\bibfnamefont {L.}~\bibnamefont {Mayer}}, \bibinfo {author} {\bibfnamefont {A.~M.}\ \bibnamefont {Brooks}}, \bibinfo {author} {\bibfnamefont {F.}~\bibnamefont {Governato}}, \ and\ \bibinfo {author} {\bibfnamefont {G.}~\bibnamefont {Lake}},\ }\href {\doibase 10.1111/j.1365-2966.2009.14757.x} {\bibfield  {journal} {\bibinfo  {journal} {Monthly Notices of the Royal Astronomical Society}\ }\textbf {\bibinfo {volume} {397}},\ \bibinfo {pages} {44–51} (\bibinfo {year} {2009})}\BibitemShut {NoStop}%
\bibitem [{\citenamefont {McCullough}\ and\ \citenamefont {Randall}(2013)}]{darkdiskMcCullough:2013jma}%
  \BibitemOpen
  \bibfield  {author} {\bibinfo {author} {\bibfnamefont {M.}~\bibnamefont {McCullough}}\ and\ \bibinfo {author} {\bibfnamefont {L.}~\bibnamefont {Randall}},\ }\href {\doibase 10.1088/1475-7516/2013/10/058} {\bibfield  {journal} {\bibinfo  {journal} {JCAP}\ }\textbf {\bibinfo {volume} {10}},\ \bibinfo {pages} {058} (\bibinfo {year} {2013})},\ \Eprint {http://arxiv.org/abs/1307.4095} {arXiv:1307.4095 [hep-ph]} \BibitemShut {NoStop}%
\bibitem [{\citenamefont {Kuhlen}\ \emph {et~al.}(2014)\citenamefont {Kuhlen}, \citenamefont {Pillepich}, \citenamefont {Guedes},\ and\ \citenamefont {Madau}}]{darkdiskKuhlen:2013tra}%
  \BibitemOpen
  \bibfield  {author} {\bibinfo {author} {\bibfnamefont {M.}~\bibnamefont {Kuhlen}}, \bibinfo {author} {\bibfnamefont {A.}~\bibnamefont {Pillepich}}, \bibinfo {author} {\bibfnamefont {J.}~\bibnamefont {Guedes}}, \ and\ \bibinfo {author} {\bibfnamefont {P.}~\bibnamefont {Madau}},\ }\href {\doibase 10.1088/0004-637X/784/2/161} {\bibfield  {journal} {\bibinfo  {journal} {Astrophys. J.}\ }\textbf {\bibinfo {volume} {784}},\ \bibinfo {pages} {161} (\bibinfo {year} {2014})},\ \Eprint {http://arxiv.org/abs/1308.1703} {arXiv:1308.1703 [astro-ph.GA]} \BibitemShut {NoStop}%
\bibitem [{\citenamefont {Schaller}\ \emph {et~al.}(2016)\citenamefont {Schaller}, \citenamefont {Frenk}, \citenamefont {Fattahi}, \citenamefont {Navarro}, \citenamefont {Oman},\ and\ \citenamefont {Sawala}}]{darkdiskSchaller:2016uot}%
  \BibitemOpen
  \bibfield  {author} {\bibinfo {author} {\bibfnamefont {M.}~\bibnamefont {Schaller}}, \bibinfo {author} {\bibfnamefont {C.~S.}\ \bibnamefont {Frenk}}, \bibinfo {author} {\bibfnamefont {A.}~\bibnamefont {Fattahi}}, \bibinfo {author} {\bibfnamefont {J.~F.}\ \bibnamefont {Navarro}}, \bibinfo {author} {\bibfnamefont {K.~A.}\ \bibnamefont {Oman}}, \ and\ \bibinfo {author} {\bibfnamefont {T.}~\bibnamefont {Sawala}},\ }\href {\doibase 10.1093/mnrasl/slw101} {\bibfield  {journal} {\bibinfo  {journal} {Mon. Not. Roy. Astron. Soc.}\ }\textbf {\bibinfo {volume} {461}},\ \bibinfo {pages} {L56} (\bibinfo {year} {2016})},\ \Eprint {http://arxiv.org/abs/1605.02770} {arXiv:1605.02770 [astro-ph.GA]} \BibitemShut {NoStop}%
\bibitem [{\citenamefont {Myeong}\ \emph {et~al.}(2018{\natexlab{b}})\citenamefont {Myeong}, \citenamefont {Evans}, \citenamefont {Belokurov}, \citenamefont {Sanders},\ and\ \citenamefont {Koposov}}]{myeong2018shardsomegacentauri}%
  \BibitemOpen
  \bibfield  {author} {\bibinfo {author} {\bibfnamefont {G.~C.}\ \bibnamefont {Myeong}}, \bibinfo {author} {\bibfnamefont {N.~W.}\ \bibnamefont {Evans}}, \bibinfo {author} {\bibfnamefont {V.}~\bibnamefont {Belokurov}}, \bibinfo {author} {\bibfnamefont {J.~L.}\ \bibnamefont {Sanders}}, \ and\ \bibinfo {author} {\bibfnamefont {S.~E.}\ \bibnamefont {Koposov}},\ }\href {https://arxiv.org/abs/1804.07050} {\enquote {\bibinfo {title} {The shards of $\omega$ centauri},}\ } (\bibinfo {year} {2018}{\natexlab{b}}),\ \Eprint {http://arxiv.org/abs/1804.07050} {arXiv:1804.07050 [astro-ph.GA]} \BibitemShut {NoStop}%
\bibitem [{\citenamefont {Petersen}\ \emph {et~al.}(2016{\natexlab{a}})\citenamefont {Petersen}, \citenamefont {Weinberg},\ and\ \citenamefont {Katz}}]{shadowbarPetersen2016}%
  \BibitemOpen
  \bibfield  {author} {\bibinfo {author} {\bibfnamefont {M.~S.}\ \bibnamefont {Petersen}}, \bibinfo {author} {\bibfnamefont {M.~D.}\ \bibnamefont {Weinberg}}, \ and\ \bibinfo {author} {\bibfnamefont {N.}~\bibnamefont {Katz}},\ }\href {\doibase 10.1093/mnras/stw2141} {\bibfield  {journal} {\bibinfo  {journal} {Monthly Notices of the Royal Astronomical Society}\ }\textbf {\bibinfo {volume} {463}},\ \bibinfo {pages} {1952–1967} (\bibinfo {year} {2016}{\natexlab{a}})}\BibitemShut {NoStop}%
\bibitem [{\citenamefont {Petersen}\ \emph {et~al.}(2016{\natexlab{b}})\citenamefont {Petersen}, \citenamefont {Katz},\ and\ \citenamefont {Weinberg}}]{shadowbarPetersen:2016vck}%
  \BibitemOpen
  \bibfield  {author} {\bibinfo {author} {\bibfnamefont {M.~S.}\ \bibnamefont {Petersen}}, \bibinfo {author} {\bibfnamefont {N.}~\bibnamefont {Katz}}, \ and\ \bibinfo {author} {\bibfnamefont {M.~D.}\ \bibnamefont {Weinberg}},\ }\href {\doibase 10.1103/PhysRevD.94.123013} {\bibfield  {journal} {\bibinfo  {journal} {Phys. Rev. D}\ }\textbf {\bibinfo {volume} {94}},\ \bibinfo {pages} {123013} (\bibinfo {year} {2016}{\natexlab{b}})},\ \Eprint {http://arxiv.org/abs/1609.01307} {arXiv:1609.01307 [hep-ph]} \BibitemShut {NoStop}%
\bibitem [{\citenamefont {Yuan}\ \emph {et~al.}(2020)\citenamefont {Yuan}, \citenamefont {Myeong}, \citenamefont {Beers}, \citenamefont {Evans}, \citenamefont {Lee}, \citenamefont {Banerjee}, \citenamefont {Gudin}, \citenamefont {Hattori}, \citenamefont {Li}, \citenamefont {Matsuno}, \citenamefont {Placco}, \citenamefont {Smith}, \citenamefont {Whitten},\ and\ \citenamefont {Zhao}}]{dynamicalrelicsYuan2020}%
  \BibitemOpen
  \bibfield  {author} {\bibinfo {author} {\bibfnamefont {Z.}~\bibnamefont {Yuan}}, \bibinfo {author} {\bibfnamefont {G.~C.}\ \bibnamefont {Myeong}}, \bibinfo {author} {\bibfnamefont {T.~C.}\ \bibnamefont {Beers}}, \bibinfo {author} {\bibfnamefont {N.~W.}\ \bibnamefont {Evans}}, \bibinfo {author} {\bibfnamefont {Y.~S.}\ \bibnamefont {Lee}}, \bibinfo {author} {\bibfnamefont {P.}~\bibnamefont {Banerjee}}, \bibinfo {author} {\bibfnamefont {D.}~\bibnamefont {Gudin}}, \bibinfo {author} {\bibfnamefont {K.}~\bibnamefont {Hattori}}, \bibinfo {author} {\bibfnamefont {H.}~\bibnamefont {Li}}, \bibinfo {author} {\bibfnamefont {T.}~\bibnamefont {Matsuno}}, \bibinfo {author} {\bibfnamefont {V.~M.}\ \bibnamefont {Placco}}, \bibinfo {author} {\bibfnamefont {M.~C.}\ \bibnamefont {Smith}}, \bibinfo {author} {\bibfnamefont {D.~D.}\ \bibnamefont {Whitten}}, \ and\ \bibinfo {author} {\bibfnamefont {G.}~\bibnamefont {Zhao}},\ }\href {\doibase 10.3847/1538-4357/ab6ef7} {\bibfield  {journal} {\bibinfo  {journal} {The Astrophysical
  Journal}\ }\textbf {\bibinfo {volume} {891}},\ \bibinfo {pages} {39} (\bibinfo {year} {2020})}\BibitemShut {NoStop}%
\bibitem [{\citenamefont {Necib}\ \emph {et~al.}(2018{\natexlab{a}})\citenamefont {Necib}, \citenamefont {Lisanti},\ and\ \citenamefont {Belokurov}}]{gaiasausage:Necib:2018iwb}%
  \BibitemOpen
  \bibfield  {author} {\bibinfo {author} {\bibfnamefont {L.}~\bibnamefont {Necib}}, \bibinfo {author} {\bibfnamefont {M.}~\bibnamefont {Lisanti}}, \ and\ \bibinfo {author} {\bibfnamefont {V.}~\bibnamefont {Belokurov}},\ }\href {\doibase 10.3847/1538-4357/ab095b} {\  (\bibinfo {year} {2018}{\natexlab{a}}),\ 10.3847/1538-4357/ab095b},\ \Eprint {http://arxiv.org/abs/1807.02519} {arXiv:1807.02519 [astro-ph.GA]} \BibitemShut {NoStop}%
\bibitem [{\citenamefont {Necib}\ \emph {et~al.}(2018{\natexlab{b}})\citenamefont {Necib}, \citenamefont {Lisanti}, \citenamefont {Garrison-Kimmel}, \citenamefont {Wetzel}, \citenamefont {Sanderson}, \citenamefont {Hopkins}, \citenamefont {Faucher-Gigu\`ere},\ and\ \citenamefont {Kere\v{s}}}]{gaiasausage:Necib:2018igl}%
  \BibitemOpen
  \bibfield  {author} {\bibinfo {author} {\bibfnamefont {L.}~\bibnamefont {Necib}}, \bibinfo {author} {\bibfnamefont {M.}~\bibnamefont {Lisanti}}, \bibinfo {author} {\bibfnamefont {S.}~\bibnamefont {Garrison-Kimmel}}, \bibinfo {author} {\bibfnamefont {A.}~\bibnamefont {Wetzel}}, \bibinfo {author} {\bibfnamefont {R.}~\bibnamefont {Sanderson}}, \bibinfo {author} {\bibfnamefont {P.~F.}\ \bibnamefont {Hopkins}}, \bibinfo {author} {\bibfnamefont {C.-A.}\ \bibnamefont {Faucher-Gigu\`ere}}, \ and\ \bibinfo {author} {\bibfnamefont {D.}~\bibnamefont {Kere\v{s}}},\ }\href {\doibase 10.3847/1538-4357/ab3afc} {\  (\bibinfo {year} {2018}{\natexlab{b}}),\ 10.3847/1538-4357/ab3afc},\ \Eprint {http://arxiv.org/abs/1810.12301} {arXiv:1810.12301 [astro-ph.GA]} \BibitemShut {NoStop}%
\bibitem [{\citenamefont {Buch}\ \emph {et~al.}(2020)\citenamefont {Buch}, \citenamefont {Fan},\ and\ \citenamefont {Leung}}]{gaiasausage:Buch:2019aiw}%
  \BibitemOpen
  \bibfield  {author} {\bibinfo {author} {\bibfnamefont {J.}~\bibnamefont {Buch}}, \bibinfo {author} {\bibfnamefont {J.}~\bibnamefont {Fan}}, \ and\ \bibinfo {author} {\bibfnamefont {J.~S.~C.}\ \bibnamefont {Leung}},\ }\href {\doibase 10.1103/PhysRevD.101.063026} {\bibfield  {journal} {\bibinfo  {journal} {Phys. Rev. D}\ }\textbf {\bibinfo {volume} {101}},\ \bibinfo {pages} {063026} (\bibinfo {year} {2020})},\ \Eprint {http://arxiv.org/abs/1910.06356} {arXiv:1910.06356 [hep-ph]} \BibitemShut {NoStop}%
\bibitem [{\citenamefont {Ibata}\ \emph {et~al.}(2001)\citenamefont {Ibata}, \citenamefont {Irwin}, \citenamefont {Lewis},\ and\ \citenamefont {Stolte}}]{streamSagittIbata:2000ys}%
  \BibitemOpen
  \bibfield  {author} {\bibinfo {author} {\bibfnamefont {R.}~\bibnamefont {Ibata}}, \bibinfo {author} {\bibfnamefont {M.}~\bibnamefont {Irwin}}, \bibinfo {author} {\bibfnamefont {G.~F.}\ \bibnamefont {Lewis}}, \ and\ \bibinfo {author} {\bibfnamefont {A.}~\bibnamefont {Stolte}},\ }\href {\doibase 10.1086/318894} {\bibfield  {journal} {\bibinfo  {journal} {Astrophys. J. Lett.}\ }\textbf {\bibinfo {volume} {547}},\ \bibinfo {pages} {L133} (\bibinfo {year} {2001})},\ \Eprint {http://arxiv.org/abs/astro-ph/0004255} {arXiv:astro-ph/0004255} \BibitemShut {NoStop}%
\bibitem [{\citenamefont {Belokurov}\ \emph {et~al.}(2007)\citenamefont {Belokurov} \emph {et~al.}}]{streamSagittBelokurov:2006kc}%
  \BibitemOpen
  \bibfield  {author} {\bibinfo {author} {\bibfnamefont {V.}~\bibnamefont {Belokurov}} \emph {et~al.},\ }\href {\doibase 10.1086/511302} {\bibfield  {journal} {\bibinfo  {journal} {Astrophys. J.}\ }\textbf {\bibinfo {volume} {658}},\ \bibinfo {pages} {337} (\bibinfo {year} {2007})},\ \Eprint {http://arxiv.org/abs/astro-ph/0605705} {arXiv:astro-ph/0605705} \BibitemShut {NoStop}%
\bibitem [{\citenamefont {Purcell}\ \emph {et~al.}(2012)\citenamefont {Purcell}, \citenamefont {Zentner},\ and\ \citenamefont {Wang}}]{streamSagittPurcell2012}%
  \BibitemOpen
  \bibfield  {author} {\bibinfo {author} {\bibfnamefont {C.~W.}\ \bibnamefont {Purcell}}, \bibinfo {author} {\bibfnamefont {A.~R.}\ \bibnamefont {Zentner}}, \ and\ \bibinfo {author} {\bibfnamefont {M.-Y.}\ \bibnamefont {Wang}},\ }\href {\doibase 10.1088/1475-7516/2012/08/027} {\bibfield  {journal} {\bibinfo  {journal} {Journal of Cosmology and Astroparticle Physics}\ }\textbf {\bibinfo {volume} {2012}},\ \bibinfo {pages} {027–027} (\bibinfo {year} {2012})}\BibitemShut {NoStop}%
\bibitem [{\citenamefont {Sollima}\ \emph {et~al.}(2014)\citenamefont {Sollima}, \citenamefont {Carretta}, \citenamefont {D’Orazi}, \citenamefont {Gratton}, \citenamefont {Bragaglia},\ and\ \citenamefont {Lucatello}}]{streamSagittSollima2014}%
  \BibitemOpen
  \bibfield  {author} {\bibinfo {author} {\bibfnamefont {A.}~\bibnamefont {Sollima}}, \bibinfo {author} {\bibfnamefont {E.}~\bibnamefont {Carretta}}, \bibinfo {author} {\bibfnamefont {V.}~\bibnamefont {D’Orazi}}, \bibinfo {author} {\bibfnamefont {R.~G.}\ \bibnamefont {Gratton}}, \bibinfo {author} {\bibfnamefont {A.}~\bibnamefont {Bragaglia}}, \ and\ \bibinfo {author} {\bibfnamefont {S.}~\bibnamefont {Lucatello}},\ }\href {\doibase 10.1093/mnras/stu1264} {\bibfield  {journal} {\bibinfo  {journal} {Monthly Notices of the Royal Astronomical Society}\ }\textbf {\bibinfo {volume} {443}},\ \bibinfo {pages} {1425–1432} (\bibinfo {year} {2014})}\BibitemShut {NoStop}%
\bibitem [{\citenamefont {O'Hare}\ and\ \citenamefont {Green}(2014)}]{OHare:2014nxd}%
  \BibitemOpen
  \bibfield  {author} {\bibinfo {author} {\bibfnamefont {C.~A.~J.}\ \bibnamefont {O'Hare}}\ and\ \bibinfo {author} {\bibfnamefont {A.~M.}\ \bibnamefont {Green}},\ }\href {\doibase 10.1103/PhysRevD.90.123511} {\bibfield  {journal} {\bibinfo  {journal} {Phys. Rev. D}\ }\textbf {\bibinfo {volume} {90}},\ \bibinfo {pages} {123511} (\bibinfo {year} {2014})},\ \Eprint {http://arxiv.org/abs/1410.2749} {arXiv:1410.2749 [astro-ph.CO]} \BibitemShut {NoStop}%
\bibitem [{\citenamefont {Kavanagh}\ and\ \citenamefont {O'Hare}(2016)}]{Kavanagh:2016xfi}%
  \BibitemOpen
  \bibfield  {author} {\bibinfo {author} {\bibfnamefont {B.~J.}\ \bibnamefont {Kavanagh}}\ and\ \bibinfo {author} {\bibfnamefont {C.~A.~J.}\ \bibnamefont {O'Hare}},\ }\href {\doibase 10.1103/PhysRevD.94.123009} {\bibfield  {journal} {\bibinfo  {journal} {Phys. Rev. D}\ }\textbf {\bibinfo {volume} {94}},\ \bibinfo {pages} {123009} (\bibinfo {year} {2016})},\ \Eprint {http://arxiv.org/abs/1609.08630} {arXiv:1609.08630 [astro-ph.CO]} \BibitemShut {NoStop}%
\bibitem [{\citenamefont {Maity}\ and\ \citenamefont {Laha}(2023)}]{Maity:2022enp}%
  \BibitemOpen
  \bibfield  {author} {\bibinfo {author} {\bibfnamefont {T.~N.}\ \bibnamefont {Maity}}\ and\ \bibinfo {author} {\bibfnamefont {R.}~\bibnamefont {Laha}},\ }\href {\doibase 10.1007/JHEP02(2023)200} {\bibfield  {journal} {\bibinfo  {journal} {JHEP}\ }\textbf {\bibinfo {volume} {02}},\ \bibinfo {pages} {200} (\bibinfo {year} {2023})},\ \Eprint {http://arxiv.org/abs/2208.14471} {arXiv:2208.14471 [hep-ph]} \BibitemShut {NoStop}%
\bibitem [{\citenamefont {{M{\"u}cket}}\ and\ \citenamefont {{Mucket}}(1989)}]{1989Ap&SS.151..177M}%
  \BibitemOpen
  \bibfield  {author} {\bibinfo {author} {\bibfnamefont {J.~P.}\ \bibnamefont {{M{\"u}cket}}}\ and\ \bibinfo {author} {\bibfnamefont {J.~P.}\ \bibnamefont {{Mucket}}},\ }\href {\doibase 10.1007/BF00643640} {\bibfield  {journal} {\bibinfo  {journal} {APSS}\ }\textbf {\bibinfo {volume} {151}},\ \bibinfo {pages} {177} (\bibinfo {year} {1989})}\BibitemShut {NoStop}%
\bibitem [{\citenamefont {Abdukerim}\ \emph {et~al.}(2024)\citenamefont {Abdukerim} \emph {et~al.}}]{PandaX-xT:2024oxq}%
  \BibitemOpen
  \bibfield  {author} {\bibinfo {author} {\bibfnamefont {A.}~\bibnamefont {Abdukerim}} \emph {et~al.} (\bibinfo {collaboration} {PandaX}),\ }\href@noop {} {\  (\bibinfo {year} {2024})},\ \Eprint {http://arxiv.org/abs/2402.03596} {arXiv:2402.03596 [hep-ex]} \BibitemShut {NoStop}%
\bibitem [{\citenamefont {Akerib}\ \emph {et~al.}(2022)\citenamefont {Akerib} \emph {et~al.}}]{SnowmassNuFog:2022ort}%
  \BibitemOpen
  \bibfield  {author} {\bibinfo {author} {\bibfnamefont {D.~S.}\ \bibnamefont {Akerib}} \emph {et~al.},\ }in\ \href@noop {} {\emph {\bibinfo {booktitle} {{Snowmass 2021}}}}\ (\bibinfo {year} {2022})\ \Eprint {http://arxiv.org/abs/2203.08084} {arXiv:2203.08084 [hep-ex]} \BibitemShut {NoStop}%
\bibitem [{\citenamefont {Acevedo}\ \emph {et~al.}(2024)\citenamefont {Acevedo}, \citenamefont {Berger},\ and\ \citenamefont {Denton}}]{Acevedo:2024wmx}%
  \BibitemOpen
  \bibfield  {author} {\bibinfo {author} {\bibfnamefont {J.~F.}\ \bibnamefont {Acevedo}}, \bibinfo {author} {\bibfnamefont {J.}~\bibnamefont {Berger}}, \ and\ \bibinfo {author} {\bibfnamefont {P.~B.}\ \bibnamefont {Denton}},\ }\href@noop {} {\  (\bibinfo {year} {2024})},\ \Eprint {http://arxiv.org/abs/2407.01670} {arXiv:2407.01670 [hep-ph]} \BibitemShut {NoStop}%
\bibitem [{\citenamefont {Ajaj}\ \emph {et~al.}(2019)\citenamefont {Ajaj} \emph {et~al.}}]{DEAP:SS:2019yzn}%
  \BibitemOpen
  \bibfield  {author} {\bibinfo {author} {\bibfnamefont {R.}~\bibnamefont {Ajaj}} \emph {et~al.} (\bibinfo {collaboration} {DEAP}),\ }\href {\doibase 10.1103/PhysRevD.100.022004} {\bibfield  {journal} {\bibinfo  {journal} {Phys. Rev. D}\ }\textbf {\bibinfo {volume} {100}},\ \bibinfo {pages} {022004} (\bibinfo {year} {2019})},\ \Eprint {http://arxiv.org/abs/1902.04048} {arXiv:1902.04048 [astro-ph.CO]} \BibitemShut {NoStop}%
\bibitem [{\citenamefont {Albanese}\ \emph {et~al.}(2021)\citenamefont {Albanese} \emph {et~al.}}]{SNOPlus:2021xpa}%
  \BibitemOpen
  \bibfield  {author} {\bibinfo {author} {\bibfnamefont {V.}~\bibnamefont {Albanese}} \emph {et~al.} (\bibinfo {collaboration} {SNO+}),\ }\href {\doibase 10.1088/1748-0221/16/08/P08059} {\bibfield  {journal} {\bibinfo  {journal} {JINST}\ }\textbf {\bibinfo {volume} {16}},\ \bibinfo {pages} {P08059} (\bibinfo {year} {2021})},\ \Eprint {http://arxiv.org/abs/2104.11687} {arXiv:2104.11687 [physics.ins-det]} \BibitemShut {NoStop}%
\bibitem [{\citenamefont {Abusleme}\ \emph {et~al.}(2022)\citenamefont {Abusleme} \emph {et~al.}}]{JUNO:2021vlw}%
  \BibitemOpen
  \bibfield  {author} {\bibinfo {author} {\bibfnamefont {A.}~\bibnamefont {Abusleme}} \emph {et~al.} (\bibinfo {collaboration} {JUNO}),\ }\href {\doibase 10.1016/j.ppnp.2021.103927} {\bibfield  {journal} {\bibinfo  {journal} {Prog. Part. Nucl. Phys.}\ }\textbf {\bibinfo {volume} {123}},\ \bibinfo {pages} {103927} (\bibinfo {year} {2022})},\ \Eprint {http://arxiv.org/abs/2104.02565} {arXiv:2104.02565 [hep-ex]} \BibitemShut {NoStop}%
\bibitem [{\citenamefont {Kouvaris}(2008)}]{Kouvaris_2008}%
  \BibitemOpen
  \bibfield  {author} {\bibinfo {author} {\bibfnamefont {C.}~\bibnamefont {Kouvaris}},\ }\href {\doibase 10.1103/physrevd.77.023006} {\bibfield  {journal} {\bibinfo  {journal} {Physical Review D}\ }\textbf {\bibinfo {volume} {77}} (\bibinfo {year} {2008}),\ 10.1103/physrevd.77.023006}\BibitemShut {NoStop}%
\bibitem [{\citenamefont {Bramante}\ \emph {et~al.}(2017)\citenamefont {Bramante}, \citenamefont {Delgado},\ and\ \citenamefont {Martin}}]{Bramante_2017}%
  \BibitemOpen
  \bibfield  {author} {\bibinfo {author} {\bibfnamefont {J.}~\bibnamefont {Bramante}}, \bibinfo {author} {\bibfnamefont {A.}~\bibnamefont {Delgado}}, \ and\ \bibinfo {author} {\bibfnamefont {A.}~\bibnamefont {Martin}},\ }\href {\doibase 10.1103/physrevd.96.063002} {\bibfield  {journal} {\bibinfo  {journal} {Physical Review D}\ }\textbf {\bibinfo {volume} {96}} (\bibinfo {year} {2017}),\ 10.1103/physrevd.96.063002}\BibitemShut {NoStop}%
\end{thebibliography}%

\end{document}